\documentclass[12pt]{article}

\usepackage{amssymb}
\usepackage{bm}
\usepackage[usenames]{color}
\usepackage{graphicx}
\usepackage{arydshln}
\usepackage{amsmath}
\usepackage{color}
\usepackage{graphics}
\usepackage{amsthm}
\usepackage{ascmac}
\usepackage{amsmath}
\usepackage{amsfonts}
\usepackage{amssymb}
\usepackage{multicol}
\usepackage{epsfig}
\usepackage{here}
\usepackage{authblk}
\usepackage{lscape}
\usepackage{url}
\usepackage{slashbox}

\usepackage{setspace}
\usepackage{listings}
\usepackage{ulem}
\makeatletter

\@addtoreset{equation}{section}
\makeatother

\newtheorem{theo}{Theorem}

\allowdisplaybreaks[4]

\newcommand{\indep}{\mathop{\perp\!\!\!\perp}}

\newcommand{\bld}{\boldsymbol}

\newcommand{\argmax}{\mathop{\rm arg~max}\limits}

\usepackage[top=1in,bottom=1in,left=1in,right=1in]{geometry}

\title{Limited-Information Maximum Likelihood based Model Selection Procedures for Binary Outcomes}
\author[1]{Shunichiro Orihara}

\affil[1]{Graduate School of Data Science, Yokohama City University, Kanagawa, Japan}

\date{}

\begin{document}
\begin{singlespace}
\maketitle
\end{singlespace}
\section*{Abstract}
Unmeasured covariates constitute one of the important problems in causal inference. Even if there are some unmeasured covariates, some instrumental variable methods such as a two-stage residual inclusion (2SRI) estimator, or a limited-information maximum likelihood (LIML) estimator can obtain an unbiased estimate for causal effects despite there being nonlinear outcomes such as binary outcomes; however, it requires that we specify not only a correct outcome model but also a correct treatment model. Therefore, detecting correct models is an important process. In this paper, we propose two model selection procedures: AIC-type and BIC-type, and confirm their properties. The proposed model selection procedures are based on a LIML estimator. We prove that a proposed BIC-type model selection procedure has model selection consistency, and confirm their properties of the proposed model selection procedures through simulation datasets.

\vspace{0.5cm}
\noindent
{\bf Keywords}: Causal inference, Consistency, Limited-information maximum lilkelihood, Model selection, Two-stage residual inclusion, Unmeasured covariates
\newpage
\section{Introduction}\noindent
Unmeasured covariates constitute one of the important problems in causal inference. When all covariates are observed, the covariates can be adjusted and an unbiased estimator for causal effects can be obtained. This is the situation of ``no unmeasured confounders" (e.g., Hern\'an and Robins, 2020). Having no unmeasured confounder is a sufficient condition to estimate an unbiased estimator for causal effects. In contrast, when some covariates are not observed (i.e., unmeasured), an unbiased estimator usually cannot be obtained. Therefore, a different sufficient assumption needs to be applied. In this paper, we focus on instrumental variable methods.

A two-stage least squares (2SLS) estimator is one of the most important two-step procedures in IV estimators for estimating causal effects when there are some unmeasured covariates (e.g., Wooldridge, 2010). A 2SLS estimators estimate causal effects in two steps. In the first step, regressing a treatment variable on instrumental variables, predictions of the treatment variable are constructed. In the second step, an outcome variable is regressed on the predicted treatment variables. A 2SLS estimator is useful, but it requires a key assumption: there is a linear relationship between an outcome variable and a treatment variable. If this assumption were violated, we might have a biased estimate for a causal effect of interest. Terza et al. (2008) introduces a two-stage residual inclusion (2SRI) estimator, similar to a control function approach (e.g., Wooldridge, 2010), which is another two-step procedure expanded to include nonlinear outcomes such as binary outcomes and multinomial outcomes. A 2SRI estimators also estimate causal effects in two steps. In the first step, regressing a treatment variable on instrumental variables; residuals of the treatment variable are constructed. In the second step, an outcome variable is regressed on not only treatment variables but also the residuals of the treatment variables. A 2SRI estimator can obtain an unbiased estimate of the causal effect even when there are nonlinear outcomes; however, we need to specify not only the correct outcome model but also the correct treatment model (Basu et al., 2017). Therefore, detecting the correct models is an important process when using 2SRI.

Information criterions are commonly used model selection procedures to select the ``correct" model. The Akaike information criterion (AIC; Akaike, 1974) is the most famous information criterion for selecting the best model in the sense of predictions of future outcomes. The Bayesian information criterion (BIC) proposed by Schwarz (1978) is another famous information criterion which has model selection consistency (i.e., it selects the correct model with probability $1$) under some assumptions (Nishii, 1984, Shao, 1997). In the field of causal inference, some reports in the literature have suggested that model selection procedures in terms of predictions such as the AIC are more appropriate (Brookhart and van der Laan, 2006, Vansteelandt et al., 2012). However, which procedures are more appropriate may depend on the situation.

As mentioned previously, it is necessary for a 2SRI estimator to specify the correct models for both a treatment variable and an outcome variable; however, a 2SRI estimator has another problem: it may derive some biased estimates as mentioned in Basu et al. (2017). To overcome this problem, we also consider a limited-information maximum likelihood (LIML) estimator (e.g.  Wooldridge, 2014), called LIMLE hereinafter in this paper. An LIMLE uses a full-likelihood approach, but it has features similar to those of a 2SRI or control function approach; an LIMLE also needs to select correct models for both a treatment variable and an outcome variable. According to the simulation results of Basu et al. (2017), a full-likelihood approach derives more accurate estimate than 2SRI. In this paper, we propose some model selection procedures for a LIMLE and confirm their properties. Since no previous studies have considered either model selection procedures for 2SRI (the control function approach) or a LIMLE combined with detecting correct models in this context, the contribution of this paper may be considered significant for these estimating procedures.

The remainder of the paper proceeds as follows. In section 2, we will present a model considered in this paper. We consider two situations: continuous treatment and dichotomous treatment; when a treatment is continuous and unmeasured covariates have bivariate normal distribution, the model is called as the Rivers-Vuong model (Rivers and Vuong, 1988). Note that LIMLEs are not limited to the situations, but for the sake of simplicity, we will only consider these situations. Also, we will derive AIC-type and BIC-type information criterions based on an LIMLE, called the LAIC and the LBIC in this paper, respectively. In this paper, the theoretical properties of the LBIC will be considered. We will prove simply that the LBIC has model selection consistency, as ordinary one. In section 3, we confirm the properties of 2SRI with model selection, LAIC, and LBIC by using simulation datasets. The simulation results show that the LAIC and LBIC have the similar properties to those of the AIC and BIC, respectively. Also, it is shown that 2SRI sometimes work when using model selection procedures in the sense that estimated causal effects have only small bias; this is the different result from Basu et al. (2017). However, an LIMLE has more accurate estimates than a 2SRI one. In the simulation, we also consider expanding the distribution of unmeasured covariates from normal distribution. Specifically, we consider that marginal distributions of unmeasured covariates have logistic distribution, and the marginal distribution obeys some copulas (e.g. Biller and Corlu, 2012, Fantazzini, 2009) when a treatment is dichotomous. Regularity conditions, calculations, derivation and proof, and supplementary information for simulations are found in the appendix.

\section{Information Criterions for the LIMLE}\noindent
Let $n$ be the sample size, and assume that $i = 1,\ 2,\dots, n$ are i.i.d. samples. $\bld{X}\in\mathbb{R}^{p}$ and $\bld{Z}\in\mathbb{R}^{K}$ denote a vector of covariates and a vector of IVs, respectively. The following relationship on unmeasured variables is assumed:
\begin{align}
\label{for1_1}
\left(
\begin{array}{c}
V\\
U
\end{array}
\right)\sim F(v,u;\xi),\ \ \left(
\begin{array}{c}
V\\
U
\end{array}
\right)\indep\left(
\begin{array}{c}
\bld{X}\\
\bld{Z}
\end{array}
\right),
\end{align}
where $\xi$is a parameter of a joint distribution $(V,U)\in \mathbb{R}^{2}$. These are the similar assumptions as in Wooldridge (2014); (\ref{for1_1}) can suggest a limited-information maximum likelihood (LIML) estimation procedure. For the sake of simplicity, we assume that $\xi$ is fixed (i.e., $\xi$ need not be estimated) in the following discussions.

Next, we introduce the models considered in this paper.
\begin{itemize}
\item Treatment model (continuous treatment)
\begin{align}
\label{for1_2}
W=\varphi_{1}(\bld{Z},\bld{X};\bld{\alpha})+V
\end{align}
\item Treatment model (dichotomous treatment)
$$
W=\bld{1}\left\{\varphi_{1}(\bld{Z},\bld{X};\bld{\alpha})+V\geq 0\right\}
$$
\item Outcome model
\begin{align}
\label{for1_3}
Y=\bld{1}\left\{\varphi_{2}(W,\bld{X};\bld{\beta})+U\geq 0\right\},
\end{align}
where $\varphi_{1}$ and $\varphi_{2}$ are two-times differentiable with respect to parameters $\bld{\alpha}$ and $\bld{\beta}$, respectively. Also, the parameter spaces of $\bld{\alpha}$ and $\bld{\beta}$ are denoted as $\Theta_{\bld{\alpha}}$ and $\Theta_{\bld{\beta}}$, respectively.
\end{itemize}
The pair of models (\ref{for1_2}) and (\ref{for1_3}) is called the Rivers-Vuong model (Rivers and Vuong, 1988), when
$$
F(v,u;\xi)=N_{2}\left(\bld{0}_{2},
\left(
\begin{array}{cc}
\sigma_{v}^2&\rho\sigma_{v}\\
\rho\sigma_{v}&1
\end{array}
\right)
\right),\ \ \xi=(\sigma_{v},\rho)^{\top},\, \sigma_{v}>0,\ \rho\neq 0.
$$
Note that the following discussions are not limited to the above treatment/outcome models but apply to other parametric models as well. However, for the sake of simplicity, we will only consider the models. To estimate parameters $\bld{\theta}=\left(\bld{\alpha}^{\top},\bld{\beta}^{\top}\right)^{\top}\in\Theta=\Theta_{\bld{\alpha}}\times\Theta_{\bld{\beta}}$, 
we consider a likelihood function $L(\bld{\theta})=\prod_{i=1}^{n}L_{i}(\bld{\theta})$ 
conditioning on $\bld{z}$ and $\bld{x}$:
\begin{align}
\label{for1_4}
L(\bld{\theta})=\prod_{i=1}^{n}f(y_{i},w_{i}|\bld{z}_{i},\bld{x}_{i};\bld{\theta})=\prod_{i=1}^{n}{\rm P}(y_{i}|w_{i},\bld{z}_{i},\bld{x}_{i};\bld{\theta})f(w_{i}|\bld{z}_{i},\bld{x}_{i};\bld{\alpha}).
\end{align}
In the following, we make explicit the specific form of the likelihood in the two cases. When Rivers-Vuong model situation, (\ref{for1_4}) becomes
\begin{align}
\label{for1_5}
L(\bld{\theta})=\prod_{i=1}^{n}\Phi\left(\frac{\varphi_{i2}(\bld{\beta})+\rho v_{i}(\bld{\alpha})}{\sqrt{1-\rho^2}}\right)^{y_{i}}\left(1-\Phi\left(\frac{\varphi_{i2}(\bld{\beta})+\rho v_{i}(\bld{\alpha})}{\sqrt{1-\rho^2}}\right)\right)^{1-y_{i}}\frac{1}{\sqrt{2\pi\sigma_{v}^2}}\exp\left\{-\frac{v_{i}^2(\bld{\alpha})}{2\sigma_{v}^2}\right\}.
\end{align}
Therefore, the log-likelihood $\ell(\bld{\theta})=\log L(\bld{\theta})$ becomes
\begin{align*}
\ell(\bld{\theta})&=\sum_{i=1}^{n}\left\{y_{i}\log\Phi\left(\frac{\varphi_{i2}(\bld{\beta})+\rho v_{i}(\bld{\alpha})}{\sqrt{1-\rho^2}}\right)+(1-y_{i})\log\left(1-\Phi\left(\frac{\varphi_{i2}(\bld{\beta})+\rho v_{i}(\bld{\alpha})}{\sqrt{1-\rho^2}}\right)\right)\right.\\
&\hspace{0.5cm}\left.-\frac{v_{i}^2(\bld{\alpha})}{2\sigma_{v}^2}-\log\left(\sqrt{2\pi\sigma_{v}^2}\right)\right\}\ \ \left(=\sum_{i=1}^{n}\ell_{i}(\bld{\theta})\right).
\end{align*}
Hereinafter, we use the following notation, as necessary:
$$
\Phi\left(\frac{\varphi_{i2}(\bld{\beta})+\rho v_{i}(\bld{\alpha})}{\sqrt{1-\rho^2}}\right)=\Phi_{i}(\bld{\theta}),\ \ \phi\left(\frac{\varphi_{i2}(\bld{\beta})+\rho v_{i}(\bld{\alpha})}{\sqrt{1-\rho^2}}\right)=\phi_{i}(\bld{\theta}).
$$
When dichotomous treatment situation, (\ref{for1_4}) becomes
\begin{align*}
L(\bld{\theta})&=\prod_{i=1}^{n}{\rm P}\left(y_{i}=1,w_{i}=1|\bld{z}_{i},\bld{x}_{i};\bld{\theta}\right)^{y_{i}w_{i}}{\rm P}\left(y_{i}=0,w_{i}=1|\bld{z}_{i},\bld{x}_{i};\bld{\theta}\right)^{(1-y_{i})w_{i}}\\
&\hspace{0.5cm}\times {\rm P}\left(y_{i}=1,w_{i}=0|\bld{z}_{i},\bld{x}_{i};\bld{\theta}\right)^{y_{i}(1-w_{i})}{\rm P}\left(y_{i}=0,w_{i}=0|\bld{z}_{i},\bld{x}_{i};\bld{\theta}\right)^{(1-y_{i})(1-w_{i})}.
\end{align*}
Therefore, the log-likelihood becomes
\begin{align}
\label{for1_5_2}
\ell(\bld{\theta})&=\sum_{i=1}^{n}\left\{y_{i}w_{i}\log\left\{1-F(\infty,-\varphi_{i2}(\bld{\beta});\xi)-F(-\varphi_{i1}(\bld{\alpha}),\infty;\xi)+F(-\varphi_{i1}(\bld{\alpha}),-\varphi_{i2}(\bld{\beta});\xi)\right\}\right.\nonumber\\
&\hspace{0.5cm}+(1-y_{i})w_{i}\log\left\{F(\infty,-\varphi_{i2}(\bld{\beta});\xi)-F(-\varphi_{i1}(\bld{\alpha}),-\varphi_{i2}(\bld{\beta});\xi)\right\}\nonumber\\
&\hspace{0.5cm}+y_{i}(1-w_{i})\log\left\{F(-\varphi_{i1}(\bld{\alpha}),\infty);\xi)-F(-\varphi_{i1}(\bld{\alpha}),-\varphi_{i2}(\bld{\beta});\xi)\right\}\nonumber\\
&\hspace{0.5cm}+\left.(1-y_{i})(1-w_{i})\log\left\{F(-\varphi_{i1}(\bld{\alpha}),-\varphi_{i2}(\bld{\beta});\xi)\right\}\right\}.
\end{align}
Note that 
$$
F(\infty,v;\xi)=\lim_{u\to\infty}F(u,v;\xi),\ \ F(u,\infty;\xi)=\lim_{v\to\infty}F(u,v;\xi).
$$
In the following simulation section, we use the above derived log-likelihoods. By maximizing the likelihood (\ref{for1_4}), a limited-information maximum likelihood estimator (LIMLE) can be derived as
$$
\hat{\bld{\theta}}=\argmax_{\bld{\theta}\in\Theta}\ell(\bld{\theta}).
$$
\subsection{Derivation of information criterions for a LIMLE and their properties}\noindent
Since a LIMLE is a likelihood-based estimator, we would like to select a model with the minimum Kullback-Leibler divergence. First, the divergence from the true distribution is\begin{align}
\label{eq9}
I\left(f(y,w|\bld{z},\bld{x}),f(y,w|\bld{z},\bld{x};\hat{\bld{\theta}})\right)&=\sum_{y}\int f(y,w|\bld{z},\bld{x})\log f(y,w|\bld{z},\bld{x})dw\nonumber\\
&\hspace{0.5cm}-\sum_{y}\int f(y,w|\bld{z},\bld{x})\log f(y,w|\bld{z},\bld{x};\hat{\bld{\theta}})dw
\end{align}
Since the first term of (\ref{eq9}) does not depend on parameter $\bld{\theta}$, the second term of (\ref{eq9}), i.e., the expected log-likelihood
$$
\sum_{y}\int f(y,w|\bld{z},\bld{x})\log f(y,w|\bld{z},\bld{x};\hat{\bld{\theta}})dw\ \ \left(=:-\frac{1}{2}d_{n}(\hat{\bld{\theta}})\right),
$$
needs to be considered. One estimator for the second term of (\ref{eq9}) is the empirical log-likelihood:
$$
\sum_{i=1}^{n}\log f(y_{i},w_{i}|\bld{z}_{i},\bld{x}_{i};\hat{\bld{\theta}})\ \ \left(=:-\frac{1}{2}\hat{D}_{n}(\hat{\bld{\theta}})\right).
$$
However, it is well known that it is biased (Konishi and Kitagawa, 1996). Thus, a bias-adjusted estimator needs to be derived; this estimator is an AIC-type information criterion. We refer to it as the LAIC:
\begin{align}
\label{eq10}
LAIC=-2\ell(\hat{\bld{\theta}})+2|\hat{\bld{\theta}}|,
\end{align}
where $\left|\cdot\right|$ gives the number of elements. The LAIC is in the same spirit as the AIC in the sense of being in terms of predictions. Also, we would also like to derive a BIC-type model selection criterion which has model selection consistency. A BIC-type model selection criterion can be constructed by changing the penalty term of (\ref{eq10}):
\begin{align}
\label{eq11}
LBIC=-2\ell(\hat{\bld{\theta}})+|\hat{\bld{\theta}}|\log(n).
\end{align}

Before the proof of the consistency of the LBIC, we define some additional notation. Let $\mathcal{A}$ and $\mathcal{B}$ be sets of model candidates for $\varphi_{i1}(\bld{\alpha})$ and $\varphi_{i2}(\bld{\beta})$, respectively. $m_{\mathcal{A}}\in\mathcal{A}$ and $m_{\mathcal{B}}\in\mathcal{B}$ denote models for $\varphi_{i1}(\bld{\alpha})$ and $\varphi_{i2}(\bld{\beta})$, and the LBIC for these models is expressed as
\begin{align*}
LBIC\left(m_{\mathcal{A}},m_{\mathcal{B}}\right)&=-2\ell\left(\hat{\bld{\alpha}}_{m_{\mathcal{A}}},\hat{\bld{\beta}}_{m_{\mathcal{B}}}\right)+\left(\left|\hat{\bld{\alpha}}_{m_{\mathcal{A}}}\right|+\left|\hat{\bld{\beta}}_{m_{\mathcal{B}}}\right|\right)\log(n)=-2\ell\left(\hat{\bld{\theta}}_{m_{\mathcal{A}},m_{\mathcal{B}}}\right)+\left|\hat{\bld{\theta}}_{m_{\mathcal{A}},m_{\mathcal{B}}}\right|\log(n).
\end{align*}
Also, let $\mathcal{A}^{c},\, \mathcal{B}^{c}$ be sets of the correct models; this is in the same spirit as Shao (1997). Therefore, $\mathcal{A}^{c}\subset\mathcal{A},\, \mathcal{B}^{c}\subset\mathcal{B}$. $m_{\mathcal{A}}^{0}\in\mathcal{A}^{c}$ and $m_{\mathcal{B}}^{0}\in\mathcal{B}^{c}$ which are the smallest pair of models in $\mathcal{A}^{c}$ and $\mathcal{B}^{c}$ in the following sense:
\begin{align}
\label{eq12}
\forall{m_{\mathcal{A}}\in\mathcal{A}^{c}\backslash \left\{m_{\mathcal{A}}^{0}\right\}},\,  \forall{m_{\mathcal{B}}\in\mathcal{B}^{c}\backslash \left\{m_{\mathcal{B}}^{0}\right\}},\ \ \left|\hat{\bld{\alpha}}_{m^{0}_{\mathcal{A}}}\right|-\left|\hat{\bld{\alpha}}_{m_{\mathcal{A}}}\right|+\left|\hat{\bld{\beta}}_{m^{0}_{\mathcal{B}}}\right|-\left|\hat{\bld{\beta}}_{m_{\mathcal{B}}}\right|<0.
\end{align}
The consistency of the LBIC can be derived in the same way as that of the BIC.
\begin{theo}{A property of the LBIC}\\
\label{theo2}
The LBIC has model selection consistency; i.e., $\forall{m_{\mathcal{A}}\in\mathcal{A}\backslash \left\{m_{\mathcal{A}}^{0}\right\}},\,  \forall{m_{\mathcal{B}}\in\mathcal{B}\backslash \left\{m_{\mathcal{B}}^{0}\right\}}$,
\begin{align}
\label{eq13}
{\rm P}\left(LBIC\left(m_{\mathcal{A}}^{0},m_{\mathcal{B}}^{0}\right)-LBIC\left(m_{\mathcal{A}},m_{\mathcal{B}}\right)<0\right)\to 1.
\end{align}
\end{theo}
\section{Simulations}\noindent
In this section, we confirm properties of our proposed model selection procedures and an ordinary 2SRI estimator with model selection. Since no previous studies have considered model selection procedures for 2SRI (or the control function approach), our simulation results may give some guides for using these estimating procedures. In the case of 2SRI, we use the ordinary AIC to select both treatment and outcome models. To confirm these properties, we summarize 1) descriptive statistics of estimates for each procedure, and 2) the number of times and proportions the true model is selected for each procedure. The number of iterations for simulations is 1,000.
\subsection{Continuous and linear treatment effect}
At first, we consider Rivers-Vuong model situation. In this setting, we confirm that our proposed model selection procedures and an ordinary 2SRI estimator with model selection work well or not. Also results of 2SLS are also summarized, as a reference result. The simulation settings are as follows:
\begin{description}
\item{{\bf Covariates:}}
$
X_{1}\sim N(0,1),\, X_{2}\sim Ber(0.5),\, X_{3}\sim N(0,1)
$
\item{{\bf An instrumental variable:}}
$
Z\sim Ber(0.5)
$
\item{{\bf Unmeasured covariates:}}
$
\left(
\begin{array}{c}
V\\
U
\end{array}
\right)\sim N\left(\bld{0}_{2},\left(
\begin{array}{cc}
1&\rho\\
&1
\end{array}
\right)\right)
$
\begin{itemize}
\item Weak correlation: $\rho=0.3$
\item Strong correlation: $\rho=0.6$
\end{itemize}
\item{{\bf A treatment model:}}
$
W=1+\alpha_{z}Z+X_{2}+X_{3}+V
$
\begin{itemize}
\item Weak instrumental variable: $\alpha_{z}=0.5$\\
$\Rightarrow$ Correlation between a treatment and an IV becomes approximately $0.15$.
\item Strong instrumental variable: $\alpha_{z}=1$\\
$\Rightarrow$ Correlation between a treatment and an IV becomes approximately $0.3$.
\end{itemize}
\item{{\bf An outcome model:}}
$
Y=\bld{1}\left\{
0.5+\beta_{w}W+0.5X_{1}+0.5X_{2}+U\geq0
\right\}
$
\begin{itemize}
\item Weak treatment effect: $\beta_{w}=0.2$\\
$\Rightarrow$ Correlation between an outcome and a treatment becomes $0.3-0.4$.
\item Strong treatment effect: $\beta_{w}=0.6$\\
$\Rightarrow$ Correlation between an outcome and a treatment becomes $0.4-0.5$.
\end{itemize}
\end{description}
To select a treatment model and an outcome model, we prepare some candidate models. Supplemental information is in the appendix. 

\begin{landscape}
\begin{figure}[h]
\begin{center}
\begin{tabular}{c}
\includegraphics[width=24cm]{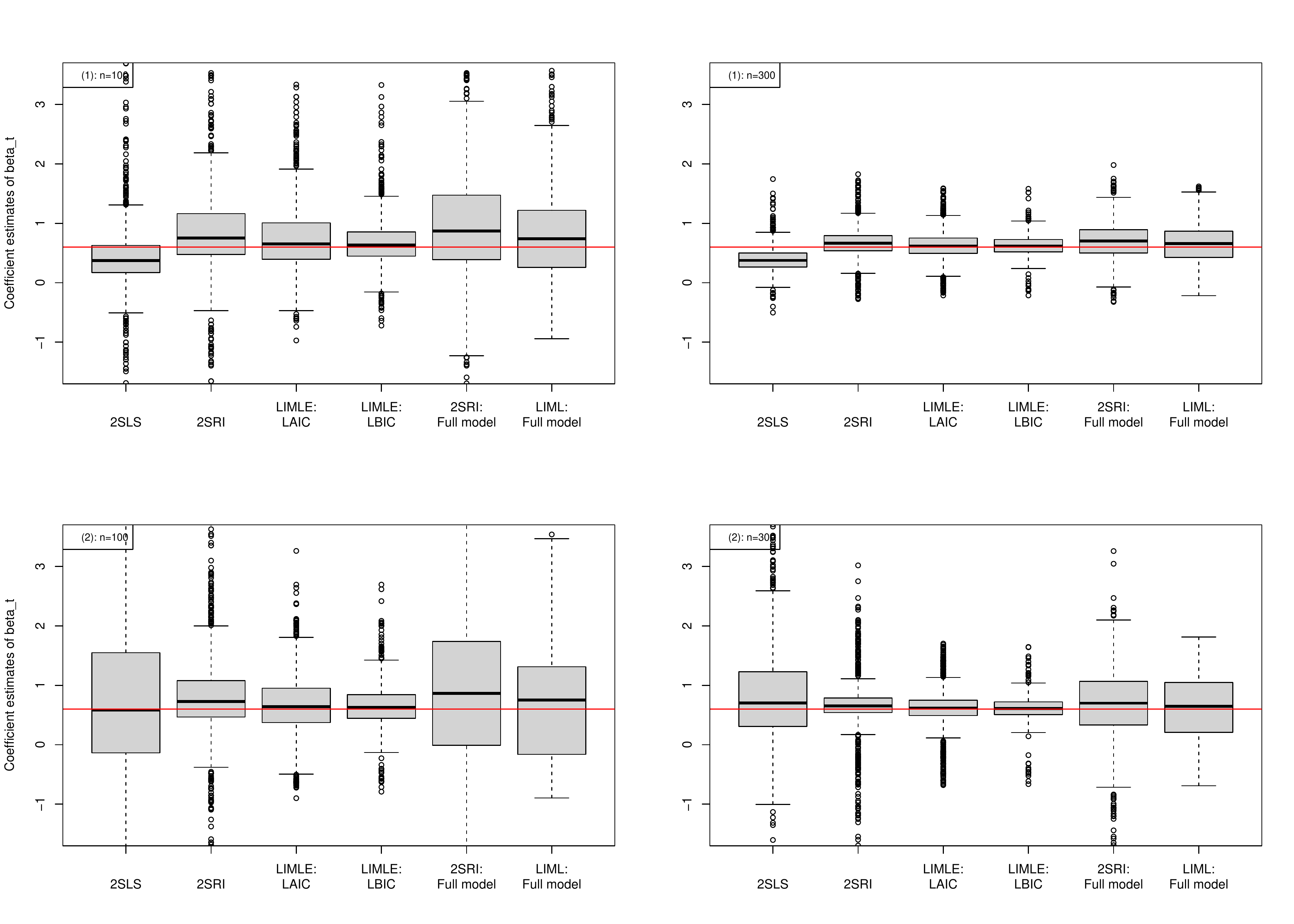}
\end{tabular}
\caption{Boxplots of descriptive statistics for each estimator (continuous and linear treatment effect) 1/2}
\label{fig1}
\end{center}
\end{figure}
\end{landscape}

\begin{landscape}
\begin{figure}[h]
\begin{center}
\begin{tabular}{c}
\includegraphics[width=24cm]{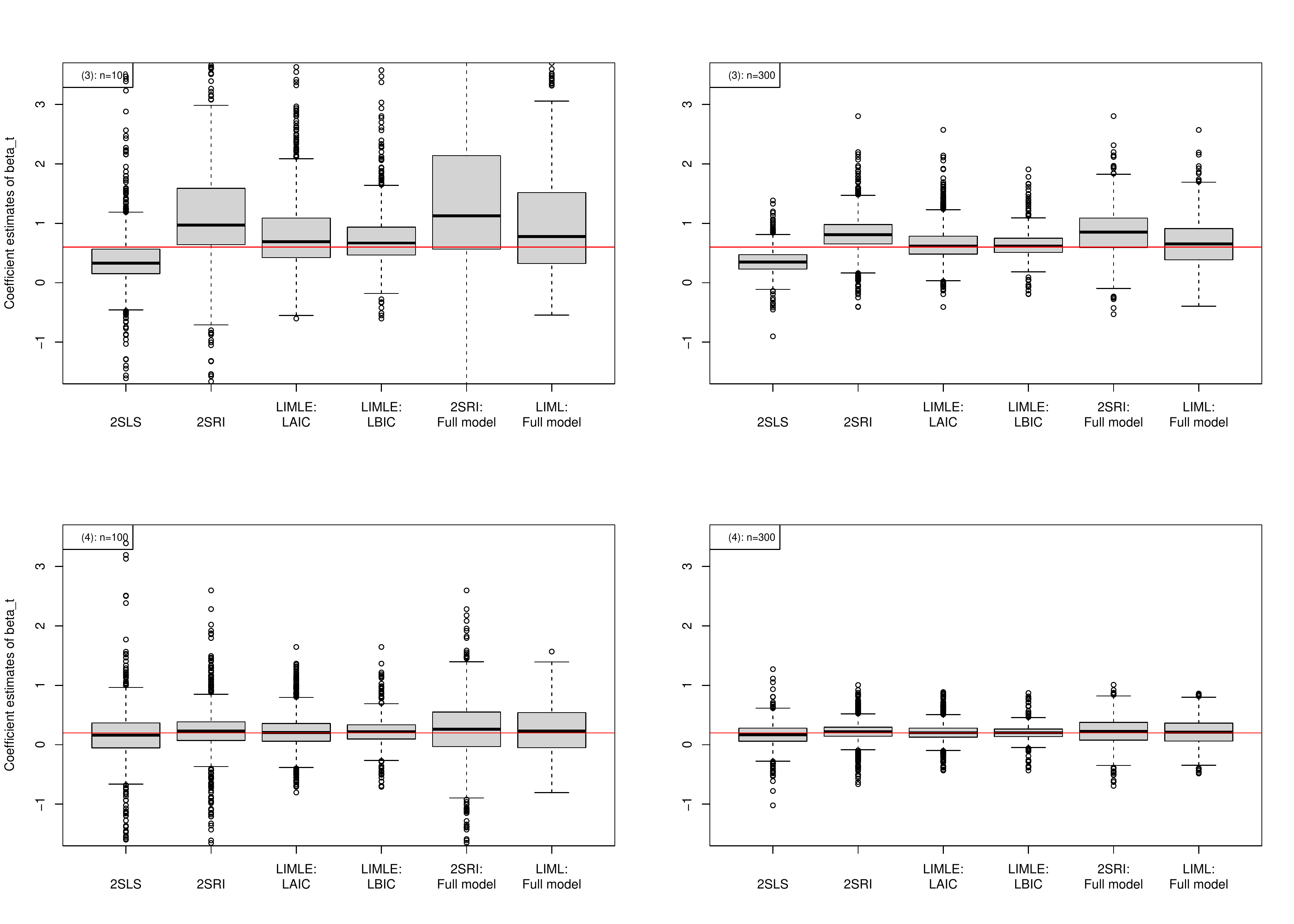}
\end{tabular}
\caption{Boxplots of descriptive statistics for each estimator (continuous and linear treatment effect) 2/2}
\label{fig2}
\end{center}
\end{figure}
\end{landscape}

Results of the simulations are summarized in tables and supplemental figures. The estimated coefficients of the treatment $W$ in the outcome model are summarized in Table \ref{tab311}, Figure \ref{fig1} and Figure \ref{fig2}, where {\bf 2SLS} is a 2SLS (without a model selection), {\bf 2SRI} is a 2SRI with a model selection, {\bf LIMLE: LAIC} and {\bf LIMLE: LBIC} are proposed model procedures, {\bf 2SRI: Full model} is a 2SRI with the largest model among the candidates, and {\bf LIMLE: Full model} is a LIMLE with the largest model among the candidates in the table, and red line is the true value in the figure. Throughout simulations, ``(1) Weak correlation, Strong IV, and Strong treatment" is the reference setting. In (1), when both the small sample ($n=100$) and large sample ($n=300$) situation, the estimator based on the LBIC is the most efficient and has the most unbiased result out of the three models with model selection procedures. The 2SRI estimator with and without model selections have large bias and low efficiency, especially, the full model result is unstable in the small sample situation. However, the both results are improved in the large sample situation. In (2), this is the weak IV situation. In this situation, LAIC and LBIC are efficient and have unbiased estimates when the large sample situation; however, there are low efficiency compared with (1). From this result, our proposed methods are also affected by the weak IV problem (Baiocchi et al., 2014), and we need to take care of the problem. The 2SRI estimators are also affected; especially, the efficiency of small sample situation is worst result out of the three situations. Thus, it may not be appropriate to use the 2SRI when there are only small sample size. In (3), our proposed methods have the worst results out of the four situations in the small sample situation: there are obvious large bias and low efficiency. However, the both results are improved in the large sample situation. The 2SRI estimator with and without model selections also have large bias and low efficiency in the small sample situation. In the large sample situation, the efficiency is improved; however, one important point is that the bias of 2SRI does not vanish. These results are the same as Basu et al. (2017). From these results, it is appeared that the residuals are not sufficiently worked to adjust unmeasured covariates when there are strong correlations between unmeasured covariates related to a treatment and an outcome. In (4), there are the similar results as (1). Almost all situations, LIMLE without model selection works well; however, it seems better to use the proposed model/variable selection procedures in the sense that the efficiency and bias of estimates. Also, 2SLS has some bias and instability compared with the other methods. Since there are no linear relationship between a treatment and an outcome, the results are natural.

The results of model selection are summarized in Table \ref{tab312}, where the column ``True model" shows how many times each method selected only the true model (i.e., the pair Model a4 and Model b2; see the appendix). The column ``Including true model" shows how many times each method selected the true or larger models (i.e., not misspecified models; see the appendix). The column ``Both true model" shows how many times a 2SRI estimator selected the true model in the first step and the second step. For selection probabilities of ``True model", LBIC does not have high probability in small sample situation; however, it is the best out of three selection procedures in all situations when large sample situation. This is the same result as {\bf Theorem \ref{theo2}}. In all situations, not only proposed methods but also 2SRI, all methods work well in the sense that selection probabilities of ``Including true model" or ``Both true model" are almost all 100\% when large sample situation.

From the above, a good choice might be considered to use LIMLE with model selection procedures, or a 2SRI with model selection procedures (except for situation (3)); however, the best model selection procedure depends on the situation, as mentioned in the Introduction. Therefore, these results are not recommendations but only illustrate the properties of each procedure. In contrast, the results of a 2SRI may not be ignored, in particular strong correlation between unmeasured covariates situations (i.e. situation (3)), since the results may include serious bias.

\newpage
\begin{landscape}
\begin{table}[h]
\caption{Summary of descriptive statistics for each estimator (coefficients of $W$, continuous and linear treatment effect)}
\scalebox{0.8}{
\begin{tabular}{|c|c||c|c|c|c||c|c|c|c|}\hline
{\bf Situation}&{\bf Method}&\multicolumn{4}{|c||}{{\bf Sample size}: n=100}&\multicolumn{4}{|c|}{{\bf Sample size}: n=300}\\\cline{3-10}
&&{Mean (SD)}&{Median (Range)}&Bias&RMSE&{Mean (SD)}&{Median (Range)}&Bias&RMSE\\\hline\hline
\begin{tabular}{c}(1) Weak\\correlation,\end{tabular}&{{\bf 2SLS}}&0.379 (3.358)&0.375 (-93.56, 23.35)&-0.221&3.366&0.396 (0.228)&0.376 (-0.50, 1.74)&-0.204&0.306\\\cline{2-10}
\begin{tabular}{c}Strong\\IV, and\end{tabular}&{{\bf 2SRI}}&4.022 (48.711)&0.751 (-174.43, 1411.34)&3.422&48.831&0.673 (0.268)&0.666 (-0.28, 1.83)&0.073&0.278\\\cline{2-10}
\begin{tabular}{c}Strong\\treatment\end{tabular}&\begin{tabular}{c}{\bf LIMLE: LAIC}\end{tabular}&1.409 (9.654)&0.652 (-0.97, 228.93)&0.809&9.688&0.625 (0.264)&0.618 (-0.21, 1.59)&0.025&0.265\\\cline{2-10}
&\begin{tabular}{c}{\bf LIMLE: LBIC}\end{tabular}&0.899 (4.402)&0.634 (-0.72, 131.56)&0.299&4.412&0.626 (0.178)&0.617 (-0.21, 1.58)&0.026&0.180\\\cline{2-10}
&\begin{tabular}{c}{\bf 2SRI: Full model}\end{tabular}&$\gg$10000 ($\gg$10000)&0.873 (-91.93, $\gg$10000)&$\gg$10000&$\gg$10000&0.703 (0.322)&0.704 (-0.32, 1.98)&0.103&0.338\\\cline{2-10}
&\begin{tabular}{c}{\bf LIMLE: Full model}\end{tabular}&0.943 (1.893)&0.739 (-0.94, 30.89)&0.343&1.924&0.651 (0.326)&0.658 (-0.22, 1.62)&0.051&0.330\\\hline\hline

\begin{tabular}{c}(2) Weak\\correlation,\end{tabular}&{{\bf 2SLS}}&-701.208 ($\gg$10000)&0.584 ($\ll$-10000, 1006.77)&-701.808&$\gg$10000&0.804 (14.203)&0.703 (-327.37, 251.92)&0.204&14.204\\\cline{2-10}
\begin{tabular}{c}Weak\\IV, and\end{tabular}&{{\bf 2SRI}}&1.203 (20.384)&0.728 (-436.7, 244.24)&0.603&20.393&0.652 (0.505)&0.655 (-2.98, 3.02)&0.052&0.508\\\cline{2-10}
\begin{tabular}{c}Strong\\treatment\end{tabular}&\begin{tabular}{c}{\bf LIMLE: LAIC}\end{tabular}&1.458 (18.760)&0.639 (-0.9, 579.24)&0.858&18.780&0.609 (0.373)&0.616 (-0.68, 1.70)&0.009&0.373\\\cline{2-10}
&\begin{tabular}{c}{\bf LIMLE: LBIC}\end{tabular}&0.839 (4.056)&0.631 (-0.79, 122.27)&0.239&4.063&0.612 (0.207)&0.611 (-0.66, 1.65)&0.012&0.207\\\cline{2-10}
&\begin{tabular}{c}{\bf 2SRI: Full model}\end{tabular}&$\gg$10000 ($\gg$10000)&0.863 (-436.70, $\gg$10000)&$\gg$10000&$\gg$10000&0.673 (0.640)&0.7 (-2.92, 3.26)&0.073&0.644\\\cline{2-10}
&\begin{tabular}{c}{\bf LIMLE: Full model}\end{tabular}&1.163 (13.241)&0.751 (-0.9, 416.97)&0.563&13.253&0.617 (0.545)&0.649 (-0.69, 1.81)&0.017&0.545\\\hline\hline

\begin{tabular}{c}(3) Strong\\correlation,\end{tabular}&{{\bf 2SLS}}&0.431 (1.263)&0.328 (-7.72, 24.63)&-0.169&1.274&0.357 (0.211)&0.348 (-0.90, 1.38)&-0.243&0.322\\\cline{2-10}
\begin{tabular}{c}Strong\\IV, and\end{tabular}&{{\bf 2SRI}}&6.968 (48.441)&0.971 (-191.22, 1154.13)&6.368&48.858&0.824 (0.334)&0.807 (-0.41, 2.80)&0.224&0.402\\\cline{2-10}
\begin{tabular}{c}Strong\\treatment\end{tabular}&\begin{tabular}{c}{\bf LIMLE: LAIC}\end{tabular}&5.588 (46.558)&0.692 (-0.60, 914.37)&4.988&46.824&0.650 (0.315)&0.618 (-0.41, 2.57)&0.050&0.319\\\cline{2-10}
&\begin{tabular}{c}{\bf LIMLE: LBIC}\end{tabular}&2.146 (23.893)&0.669 (-0.60, 672.81)&1.546&23.943&0.634 (0.216)&0.617 (-0.19, 1.91)&0.034&0.219\\\cline{2-10}
&\begin{tabular}{c}{\bf 2SRI: Full model}\end{tabular}&$\gg$10000 ($\gg$10000)&1.126 (-191.22, $\gg$10000)&$\gg$10000&$\gg$10000&0.848 (0.391)&0.853 (-0.53, 2.80)&0.248&0.463\\\cline{2-10}
&\begin{tabular}{c}{\bf LIMLE: Full model}\end{tabular}&3.043 (16.018)&0.779 (-0.54, 289.51)&2.443&16.203&0.675 (0.384)&0.654 (-0.39, 2.57)&0.075&0.391\\\hline\hline

\begin{tabular}{c}(4) Weak\\correlation,\end{tabular}&{{\bf 2SLS}}&0.113 (1.053)&0.162 (-14.76, 9.38)&-0.087&1.057&0.163 (0.206)&0.168 (-1.02, 1.27)&-0.037&0.209\\\cline{2-10}
\begin{tabular}{c}Strong\\IV, and\end{tabular}&{{\bf 2SRI}}&0.230 (0.427)&0.231 (-2.08, 2.60)&0.030&0.428&0.219 (0.187)&0.216 (-0.66, 1.01)&0.019&0.188\\\cline{2-10}
\begin{tabular}{c}Weak\\treatment\end{tabular}&\begin{tabular}{c}{\bf LIMLE: LAIC}\end{tabular}&0.223 (0.325)&0.206 (-0.80, 1.65)&0.023&0.326&0.208 (0.172)&0.203 (-0.43, 0.89)&0.008&0.172\\\cline{2-10}
&\begin{tabular}{c}{\bf LIMLE: LBIC}\end{tabular}&0.223 (0.233)&0.217 (-0.71, 1.65)&0.023&0.234&0.205 (0.124)&0.202 (-0.43, 0.87)&0.005&0.124\\\cline{2-10}
&\begin{tabular}{c}{\bf 2SRI: Full model}\end{tabular}&0.247 (0.524)&0.260 (-2.26, 2.60)&0.047&0.526&0.223 (0.230)&0.225 (-0.69, 1.01)&0.023&0.231\\\cline{2-10}
&\begin{tabular}{c}{\bf LIMLE: Full model}\end{tabular}&0.241 (0.420)&0.230 (-0.80, 1.57)&0.041&0.422&0.214 (0.220)&0.209 (-0.48, 0.86)&0.014&0.220\\\hline
\end{tabular}
}
\label{tab311}
\end{table}
\end{landscape}
\newpage
\begin{landscape}
\begin{table}[h]
\begin{center}
\caption{Summary of the results of model selection for each estimator (continuous and linear treatment effect)}
\scalebox{0.8}{
\begin{tabular}{|c|c|c||c|c|c||c|c|c|}\hline
{{\bf Situation}}&{{\bf Method}}&Step&\multicolumn{3}{|c||}{{\bf Sample size}: n=100}&\multicolumn{3}{|c|}{{\bf Sample size}: n=300}\\\cline{4-9}
&&&
\begin{tabular}{c}
True model\\
n (\%)
\end{tabular}&
\begin{tabular}{c}
Including true model\\
n (\%)
\end{tabular}&
\begin{tabular}{c}
Both true model\\
n (\%)
\end{tabular}&
\begin{tabular}{c}
True model\\
n (\%)
\end{tabular}&
\begin{tabular}{c}
Including true model\\
n (\%)
\end{tabular}&
\begin{tabular}{c}
Both true model\\
n (\%)
\end{tabular}\\\hline\hline
\begin{tabular}{c}(1) Weak\\correlation,\end{tabular}&{{\bf 2SRI}}&1st&871 (87.1)&1000 (100)&840 (84.0)&872 (87.2)&1000 (100)&1000 (100)\\\cline{3-5}\cline{7-8}
\begin{tabular}{c}Strong\\IV, and\end{tabular}&&2nd&450 (45.0)&840 (84.0)&&672 (67.2)&1000 (100)&\\\cline{2-9}
\begin{tabular}{c}Strong\\treatment\end{tabular}&\begin{tabular}{c}{\bf LIMLE: LAIC}\end{tabular}&-&390 (39.0)&837 (83.7)&-&588 (58.8)&1000 (100)&-\\\cline{2-9}
&\begin{tabular}{c}{\bf LIMLE: LBIC}\end{tabular}&-&347 (34.7)&430 (43.0)&-&846 (84.6)&880 (88.0)&-\\\hline\hline
\begin{tabular}{c}(2) Weak\\correlation,\end{tabular}&{{\bf 2SRI}}&1st&871 (87.1)&1000 (100)&867 (86.7)&872 (87.2)&1000 (100)&998 (99.8)\\\cline{3-5}\cline{7-8}
\begin{tabular}{c}Weak\\IV, and\end{tabular}&&2nd&485 (48.5)&867 (86.7)&&668 (66.8)&998 (99.8)&\\\cline{2-9}
\begin{tabular}{c}Strong\\treatment\end{tabular}&\begin{tabular}{c}{\bf LIMLE: LAIC}\end{tabular}&-&422 (42.2)&866 (86.6)&-&593 (59.3)&998 (99.8)&-\\\cline{2-9}
&\begin{tabular}{c}{\bf LIMLE: LBIC}\end{tabular}&-&426 (42.6)&494 (49.4)&-&888 (88.8)&923 (92.3)&-\\\hline\hline
\begin{tabular}{c}(3) Strong\\correlation,\end{tabular}&{{\bf 2SRI}}&1st&881 (88.1)&1000 (100)&894 (89.4)&876 (87.6)&1000 (100)&999 (99.9)\\\cline{3-5}\cline{7-8}
\begin{tabular}{c}Strong\\IV, and\end{tabular}&&2nd&493 (49.3)&894 (89.4)&&617 (61.7)&999 (99.9)&\\\cline{2-9}
\begin{tabular}{c}Strong\\treatment\end{tabular}&\begin{tabular}{c}{\bf LIMLE: LAIC}\end{tabular}&-&452 (45.2)&887 (88.7)&-&553 (55.3)&999 (99.9)&-\\\cline{2-9}
&\begin{tabular}{c}{\bf LIMLE: LBIC}\end{tabular}&-&451 (45.1)&518 (51.8)&-&900 (90.0)&951 (95.1)&-\\\hline\hline
\begin{tabular}{c}(4) Weak\\correlation,\end{tabular}&{{\bf 2SRI}}&1st&871 (87.1)&1000 (100)&944 (94.4)&872 (87.2)&1000 (100)&1000 (100)\\\cline{3-5}\cline{7-8}
\begin{tabular}{c}Strong\\IV, and\end{tabular}&&2nd&577 (57.7)&944 (94.4)&&677 (67.7)&1000 (100)&\\\cline{2-9}
\begin{tabular}{c}Weak\\treatment\end{tabular}&\begin{tabular}{c}{\bf LIMLE: LAIC}\end{tabular}&-&511 (51.1)&942 (94.2)&-&609 (60.9)&1000 (100)&-\\\cline{2-9}
&\begin{tabular}{c}{\bf LIMLE: LBIC}\end{tabular}&-&595 (59.5)&674 (67.4)&-&953 (95.3)&993 (99.3)&-\\\hline
\end{tabular}
}
\label{tab312}
\end{center}
\end{table}
\end{landscape}
\newpage
\subsection{Dichotomous and heterogeneous treatment effect}
Next, we consider dichotomous and heterogeneous treatment effect, and normal/non-normal unmeasured covariates situation. When there are non-normal unmeasured covariates, to consider joint distributions of unmeasured covariates, we use copula-based joint distributions (e.g. Biller and Corlu, 2012, Fantazzini, 2009). Except for the following update, the simulation settings are the same as previous one:
\begin{description}
\item{{\bf Unmeasured covariates (Normal):}}
$
\left(
\begin{array}{c}
V\\
U
\end{array}
\right)\sim N\left(\bld{0}_{2},\left(
\begin{array}{cc}
1&\rho\\
&1
\end{array}
\right)\right)
$
\begin{itemize}
\item Weak correlation: $\rho=0.3$
\item Strong correlation: $\rho=0.6$
\end{itemize}
\item{{\bf Unmeasured covariates (Non-normal):}}
$V\sim Logit(0,1),\ U\sim Logit(0,1)$
\begin{itemize}
\item Heavy-tailed joint distribution: t-copula
\item Asymmetric joint distribution: Clayton-copula
\end{itemize}
Note that to decide a correlation between $V$ and $U$ as the following settings approximately, we decide a parameter of the copulas which relates to the strength of marginal distributions' relationship.
\begin{itemize}
\item Weak correlation: $\approx0.3$
\item Strong correlation: $\approx0.6$
\end{itemize}
\item{{\bf A treatment model:}}
$
W=\bld{1}\left\{\alpha_{0}+\alpha_{z}Z+X_{2}+X_{3}+V\geq0
\right\}
$
\begin{itemize}
\item Weak instrumental variable: $\alpha_{0}=0.05,\, \alpha_{z}=0.6$\\
$\Rightarrow$ Correlation between a treatment and an IV becomes approximately $0.15$.
\item Strong instrumental variable: $\alpha_{0}=-0.2,\, \alpha_{z}=1.2$\\
$\Rightarrow$ Correlation between a treatment and an IV becomes approximately $0.3$.
\end{itemize}
\item{{\bf An outcome model:}}
$
Y=\bld{1}\left\{
-0.2+\beta_{w1}W-(W\times X_{1})+0.5X_{1}+0.5X_{2}+U\geq0
\right\}
$
\begin{itemize}
\item Weak treatment effect: $\beta_{w1}=0.5$\\
$\Rightarrow$ Correlation between an outcome and a treatment becomes approximately $0.20$.
\item Strong treatment effect: $\beta_{w1}=1.5$\\
$\Rightarrow$ Correlation between an outcome and a treatment becomes approximately $0.48$.
\end{itemize}
\end{description}
In this simulation, the selection of an outcome model is only considered. To select an outcome model, we prepare some candidate models. Supplemental information is in the appendix. 

\subsubsection{Specified the correct joint distribution of unmeasured covariates}
We consider that there are normal unmeasured covariates as the specified correct joint distribution. Therefore, the correct model can be specified in the candidate models. Under this setting, we confirm that our proposed model selection procedures and an ordinary 2SRI estimator with model selection work well or not, nevertheless there are more complicated situation than the previous simulation setting.

\begin{landscape}
\begin{figure}[h]
\begin{center}
\begin{tabular}{c}
\includegraphics[width=24cm]{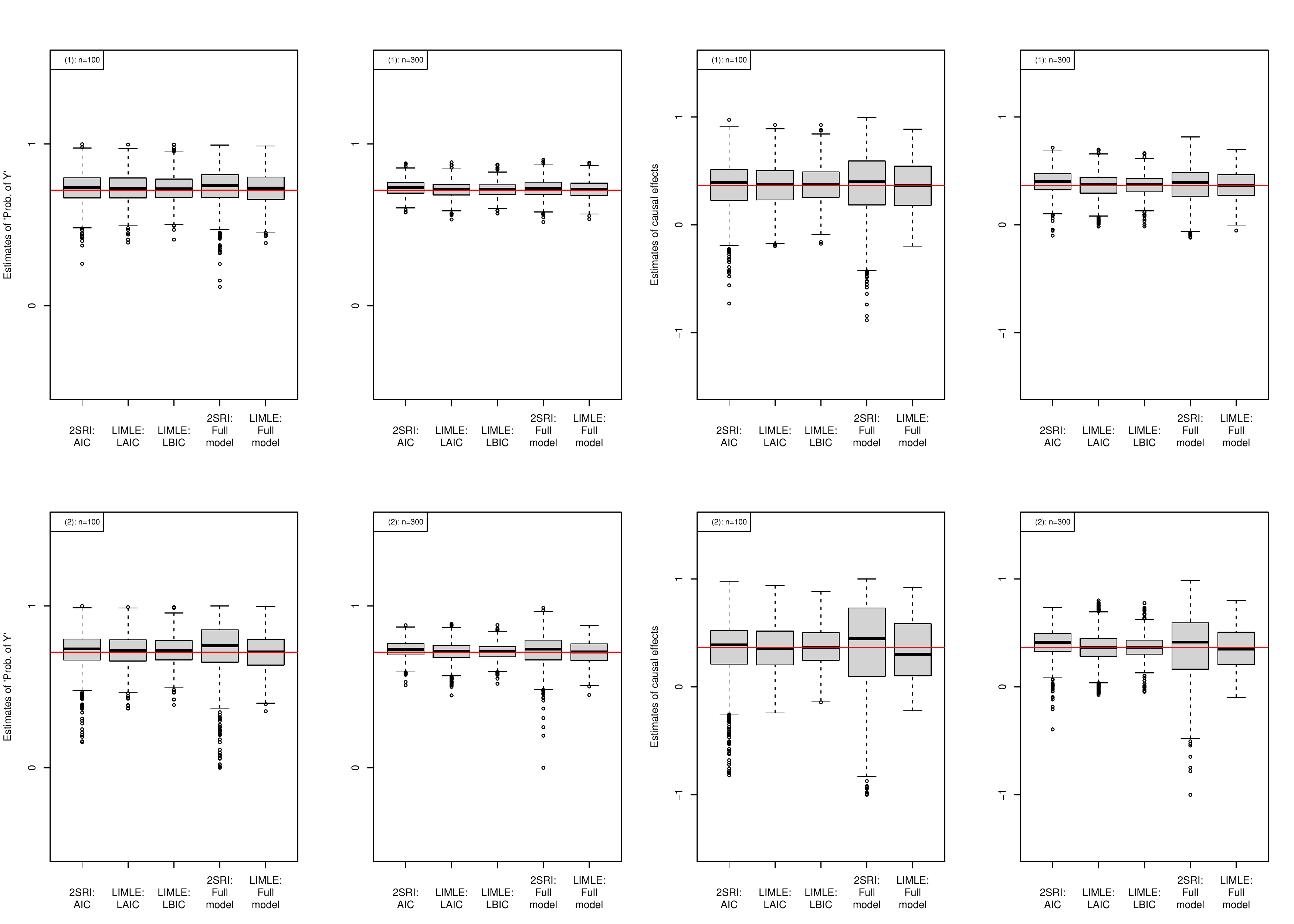}
\end{tabular}
\caption{Boxplots of descriptive statistics for each estimator (dichotomous and heterogeneous, normal dist.) 1/2}
\label{fig3}
\end{center}
\end{figure}
\end{landscape}

\begin{landscape}
\begin{figure}[h]
\begin{center}
\begin{tabular}{c}
\includegraphics[width=24cm]{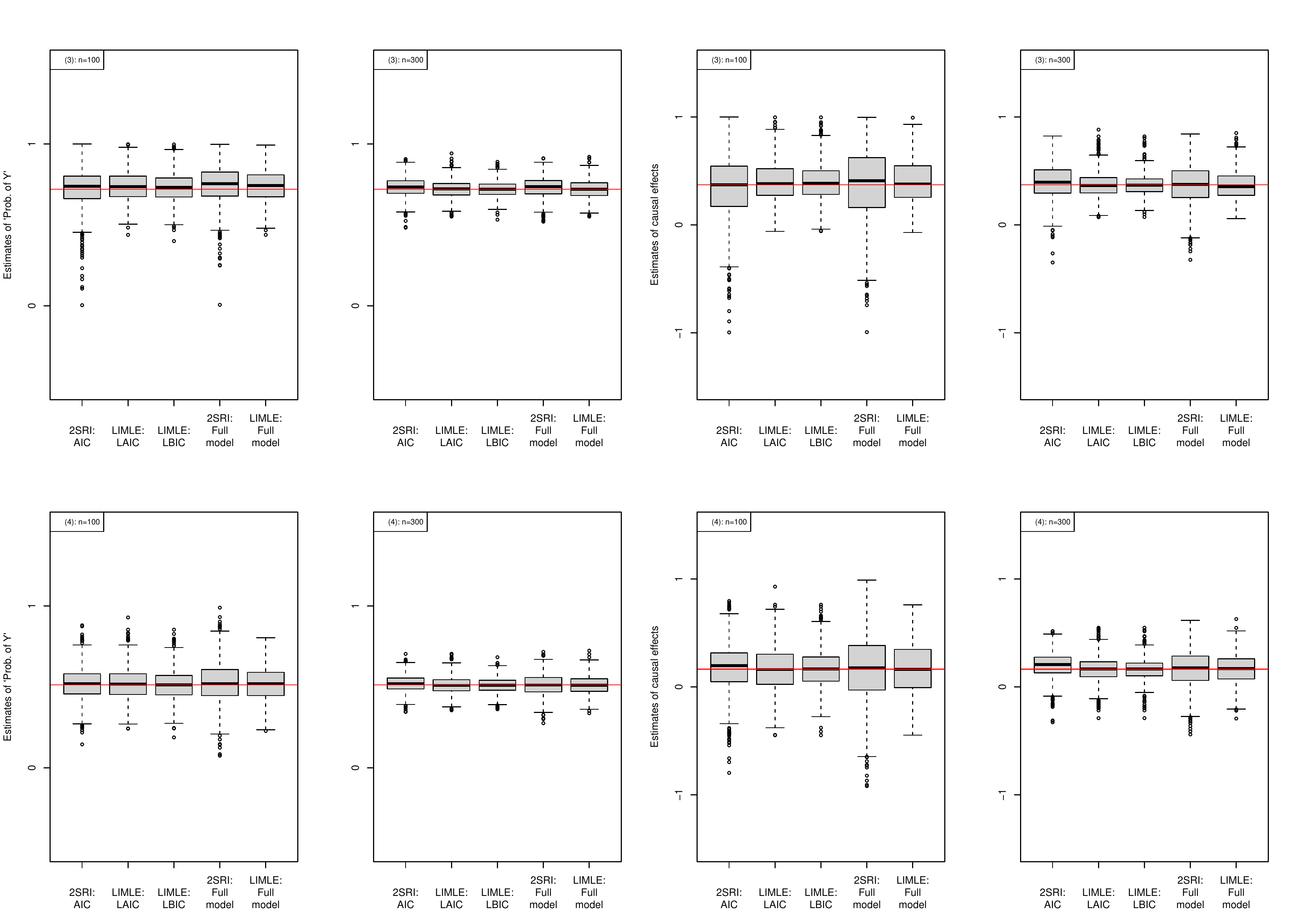}
\end{tabular}
\caption{Boxplots of descriptive statistics for each estimator (dichotomous and heterogeneous, normal dist.) 2/2}
\label{fig4}
\end{center}
\end{figure}
\end{landscape}

Results of the simulations are summarized in tables and supplemental figures. The estimated probability of $Y_{1}$ and causal effects are summarized in Table \ref{tab321}, Table \ref{tab322}, Figure \ref{fig3}, and Figure \ref{fig4}. Also, results of the model selection is summarized in Table \ref{tab323}. Both the estimates and model selection, all simulation results are similar as the previous one. Also, our proposed procedures have small bias and low efficiency compared with the 2SRI when there are only small sample size. Note that compared to the proposed method, the results of the 2SRI have some bias, however these magnitude is not so large; this is the different point from the previous simulation. Although the 2SRI does not assume a dichotomous treatment situation, the results of simulation show that the 2SRI works to some extent.

\newpage
\begin{landscape}
\begin{table}[h]
\caption{Summary of causal effects for each estimator (prob. of $Y_{1}$, dichotomous and heterogeneous, normal dist.)}
\scalebox{0.8}{
\begin{tabular}{|c|c||c|c|c|c||c|c|c|c|}\hline
{\bf Situation}&{\bf Method}&\multicolumn{4}{|c||}{{\bf Sample size}: n=100}&\multicolumn{4}{|c|}{{\bf Sample size}: n=300}\\\cline{3-10}
&&{Mean (SD)}&{Median (Range)}&Bias&RMSE&{Mean (SD)}&{Median (Range)}&Bias&RMSE\\\hline\hline
\begin{tabular}{c}(1) Weak\\correlation,\end{tabular}&{{\bf 2SRI}}&0.727 (0.096)&0.731 (0.26, 1.00)&0.012&0.097&0.728 (0.048)&0.730 (0.58, 0.88)&0.013&0.049\\\cline{2-10}
\begin{tabular}{c}Strong\\IV, and\end{tabular}&\begin{tabular}{c}{\bf LIMLE: LAIC}\end{tabular}&0.726 (0.091)&0.725 (0.39, 1.00)&0.012&0.092&0.718 (0.049)&0.719 (0.53, 0.89)&0.003&0.049\\\cline{2-10}
\begin{tabular}{c}Strong\\treatment\end{tabular}&\begin{tabular}{c}{\bf LIMLE: LBIC}\end{tabular}&0.726 (0.085)&0.724 (0.41, 1.00)&0.011&0.086&0.718 (0.044)&0.719 (0.57, 0.87)&0.004&0.044\\\cline{2-10}
&\begin{tabular}{c}{\bf 2SRI: Full model}\end{tabular}&0.736 (0.114)&0.742 (0.12, 0.99)&0.021&0.116&0.724 (0.058)&0.725 (0.52, 0.90)&0.010&0.058\\\cline{2-10}
&\begin{tabular}{c}{\bf LIMLE: Full model}\end{tabular}&0.726 (0.097)&0.727 (0.39, 0.99)&0.012&0.097&0.718 (0.055)&0.721 (0.54, 0.88)&0.003&0.055\\\hline\hline

\begin{tabular}{c}(2) Weak\\correlation,\end{tabular}&{{\bf 2SRI}}&0.723 (0.115)&0.735 (0.16, 1.00)&0.008&0.115&0.732 (0.053)&0.732 (0.51, 0.88)&0.017&0.056\\\cline{2-10}
\begin{tabular}{c}Weak\\IV, and\end{tabular}&\begin{tabular}{c}{\bf LIMLE: LAIC}\end{tabular}&0.722 (0.099)&0.726 (0.37, 0.99)&0.008&0.100&0.715 (0.064)&0.720 (0.45, 0.89)&0.001&0.064\\\cline{2-10}
\begin{tabular}{c}Strong\\treatment\end{tabular}&\begin{tabular}{c}{\bf LIMLE: LBIC}\end{tabular}&0.725 (0.087)&0.726 (0.39, 0.99)&0.010&0.087&0.718 (0.047)&0.718 (0.52, 0.88)&0.003&0.048\\\cline{2-10}
&\begin{tabular}{c}{\bf 2SRI: Full model}\end{tabular}&0.734 (0.168)&0.755 (0.00, 1.00)&0.019&0.169&0.722 (0.100)&0.733 (0.00, 0.99)&0.008&0.100\\\cline{2-10}
&\begin{tabular}{c}{\bf LIMLE: Full model}\end{tabular}&0.716 (0.110)&0.717 (0.35, 1.00)&0.001&0.110&0.712 (0.074)&0.716 (0.45, 0.88)&-0.003&0.074\\\hline\hline

\begin{tabular}{c}(3) Strong\\correlation,\end{tabular}&{{\bf 2SRI}}&0.724 (0.125)&0.738 (0.00, 1.00)&0.003&0.125&0.733 (0.059)&0.734 (0.48, 0.91)&0.012&0.061\\\cline{2-10}
\begin{tabular}{c}Strong\\IV, and\end{tabular}&\begin{tabular}{c}{\bf LIMLE: LAIC}\end{tabular}&0.739 (0.094)&0.737 (0.44, 1.00)&0.019&0.096&0.721 (0.053)&0.723 (0.55, 0.94)&0.001&0.053\\\cline{2-10}
\begin{tabular}{c}Strong\\treatment\end{tabular}&\begin{tabular}{c}{\bf LIMLE: LBIC}\end{tabular}&0.732 (0.091)&0.731 (0.40, 1.00)&0.012&0.092&0.721 (0.048)&0.721 (0.53, 0.89)&0.000&0.048\\\cline{2-10}
&\begin{tabular}{c}{\bf 2SRI: Full model}\end{tabular}&0.744 (0.120)&0.754 (0.01, 1.00)&0.023&0.122&0.732 (0.062)&0.736 (0.52, 0.91)&0.012&0.064\\\cline{2-10}
&\begin{tabular}{c}{\bf LIMLE: Full model}\end{tabular}&0.742 (0.096)&0.742 (0.44, 0.99)&0.022&0.099&0.721 (0.058)&0.721 (0.55, 0.92)&0.000&0.058\\\hline\hline

\begin{tabular}{c}(4) Weak\\correlation,\end{tabular}&{{\bf 2SRI}}&0.519 (0.101)&0.520 (0.14, 0.88)&0.007&0.101&0.522 (0.052)&0.521 (0.35, 0.70)&0.009&0.053\\\cline{2-10}
\begin{tabular}{c}Strong\\IV, and\end{tabular}&\begin{tabular}{c}{\bf LIMLE: LAIC}\end{tabular}&0.518 (0.098)&0.517 (0.24, 0.93)&0.006&0.098&0.511 (0.053)&0.509 (0.36, 0.70)&-0.002&0.053\\\cline{2-10}
\begin{tabular}{c}Weak\\treatment\end{tabular}&\begin{tabular}{c}{\bf LIMLE: LBIC}\end{tabular}&0.512 (0.091)&0.512 (0.19, 0.85)&0.000&0.091&0.510 (0.047)&0.510 (0.36, 0.68)&-0.003&0.047\\\cline{2-10}
&\begin{tabular}{c}{\bf 2SRI: Full model}\end{tabular}&0.522 (0.129)&0.521 (0.08, 0.99)&0.009&0.129&0.514 (0.065)&0.512 (0.28, 0.72)&0.001&0.065\\\cline{2-10}
&\begin{tabular}{c}{\bf LIMLE: Full model}\end{tabular}&0.520 (0.103)&0.519 (0.23, 0.80)&0.007&0.103&0.512 (0.058)&0.510 (0.34, 0.72)&-0.001&0.058\\\hline
\end{tabular}
}
\label{tab321}
\end{table}
\end{landscape}

\newpage
\begin{landscape}
\begin{table}[h]
\caption{Summary of causal effects for each estimator (causal effect, dichotomous and heterogeneous, normal dist.)}
\scalebox{0.8}{
\begin{tabular}{|c|c||c|c|c|c||c|c|c|c|}\hline
{\bf Situation}&{\bf Method}&\multicolumn{4}{|c||}{{\bf Sample size}: n=100}&\multicolumn{4}{|c|}{{\bf Sample size}: n=300}\\\cline{3-10}
&&{Mean (SD)}&{Median (Range)}&Bias&RMSE&{Mean (SD)}&{Median (Range)}&Bias&RMSE\\\hline\hline
\begin{tabular}{c}(1) Weak\\correlation,\end{tabular}&{{\bf 2SRI}}&0.355 (0.239)&0.391 (-0.73, 0.97)&-0.010&0.239&0.397 (0.115)&0.401 (-0.10, 0.71)&0.031&0.119\\\cline{2-10}
\begin{tabular}{c}Strong\\IV, and\end{tabular}&\begin{tabular}{c}{\bf LIMLE: LAIC}\end{tabular}&0.370 (0.205)&0.374 (-0.20, 0.93)&0.004&0.205&0.367 (0.119)&0.371 (-0.02, 0.70)&0.002&0.119\\\cline{2-10}
\begin{tabular}{c}Strong\\treatment\end{tabular}&\begin{tabular}{c}{\bf LIMLE: LBIC}\end{tabular}&0.374 (0.173)&0.373 (-0.17, 0.93)&0.008&0.173&0.369 (0.092)&0.370 (-0.01, 0.66)&0.003&0.093\\\cline{2-10}
&\begin{tabular}{c}{\bf 2SRI: Full model}\end{tabular}&0.369 (0.308)&0.400 (-0.88, 0.99)&0.003&0.308&0.372 (0.166)&0.392 (-0.12, 0.81)&0.006&0.166\\\cline{2-10}
&\begin{tabular}{c}{\bf LIMLE: Full model}\end{tabular}&0.366 (0.230)&0.365 (-0.20, 0.89)&0.000&0.230&0.364 (0.137)&0.369 (-0.05, 0.70)&-0.001&0.137\\\hline\hline

\begin{tabular}{c}(2) Weak\\correlation,\end{tabular}&{{\bf 2SRI}}&0.347 (0.287)&0.390 (-0.82, 0.97)&-0.018&0.288&0.407 (0.141)&0.412 (-0.39, 0.73)&0.041&0.147\\\cline{2-10}
\begin{tabular}{c}Weak\\IV, and\end{tabular}&\begin{tabular}{c}{\bf LIMLE: LAIC}\end{tabular}&0.359 (0.228)&0.357 (-0.24, 0.94)&-0.007&0.228&0.361 (0.166)&0.365 (-0.07, 0.80)&-0.004&0.166\\\cline{2-10}
\begin{tabular}{c}Strong\\treatment\end{tabular}&\begin{tabular}{c}{\bf LIMLE: LBIC}\end{tabular}&0.372 (0.178)&0.369 (-0.14, 0.88)&0.006&0.178&0.367 (0.106)&0.367 (-0.05, 0.78)&0.001&0.106\\\cline{2-10}
&\begin{tabular}{c}{\bf 2SRI: Full model}\end{tabular}&0.368 (0.444)&0.446 (-1.00, 1.00)&0.002&0.444&0.362 (0.304)&0.414 (-1.00, 0.99)&-0.004&0.304\\\cline{2-10}
&\begin{tabular}{c}{\bf LIMLE: Full model}\end{tabular}&0.336 (0.275)&0.303 (-0.22, 0.92)&-0.029&0.277&0.351 (0.202)&0.351 (-0.10, 0.80)&-0.014&0.202\\\hline\hline

\begin{tabular}{c}(3) Strong\\correlation,\end{tabular}&{{\bf 2SRI}}&0.338 (0.290)&0.370 (-1.00, 1.00)&-0.032&0.292&0.395 (0.159)&0.396 (-0.35, 0.82)&0.024&0.161\\\cline{2-10}
\begin{tabular}{c}Strong\\IV, and\end{tabular}&\begin{tabular}{c}{\bf LIMLE: LAIC}\end{tabular}&0.406 (0.188)&0.383 (-0.06, 1.00)&0.035&0.192&0.372 (0.117)&0.366 (0.07, 0.88)&0.001&0.117\\\cline{2-10}
\begin{tabular}{c}Strong\\treatment\end{tabular}&\begin{tabular}{c}{\bf LIMLE: LBIC}\end{tabular}&0.399 (0.168)&0.384 (-0.06, 1.00)&0.028&0.171&0.371 (0.094)&0.367 (0.07, 0.82)&0.000&0.094\\\cline{2-10}
&\begin{tabular}{c}{\bf 2SRI: Full model}\end{tabular}&0.368 (0.325)&0.410 (-0.99, 1.00)&-0.003&0.325&0.367 (0.184)&0.375 (-0.32, 0.84)&-0.003&0.184\\\cline{2-10}
&\begin{tabular}{c}{\bf LIMLE: Full model}\end{tabular}&0.407 (0.204)&0.380 (-0.07, 0.99)&0.036&0.207&0.369 (0.135)&0.354 (0.06, 0.85)&-0.002&0.135\\\hline\hline

\begin{tabular}{c}(4) Weak\\correlation,\end{tabular}&{{\bf 2SRI}}&0.172 (0.223)&0.196 (-0.80, 0.80)&0.008&0.223&0.199 (0.115)&0.206 (-0.33, 0.52)&0.035&0.121\\\cline{2-10}
\begin{tabular}{c}Strong\\IV, and\end{tabular}&\begin{tabular}{c}{\bf LIMLE: LAIC}\end{tabular}&0.168 (0.207)&0.161 (-0.45, 0.93)&0.004&0.207&0.163 (0.121)&0.166 (-0.29, 0.55)&-0.001&0.121\\\cline{2-10}
\begin{tabular}{c}Weak\\treatment\end{tabular}&\begin{tabular}{c}{\bf LIMLE: LBIC}\end{tabular}&0.169 (0.173)&0.165 (-0.45, 0.76)&0.005&0.173&0.162 (0.096)&0.166 (-0.29, 0.55)&-0.002&0.096\\\cline{2-10}
&\begin{tabular}{c}{\bf 2SRI: Full model}\end{tabular}&0.161 (0.317)&0.178 (-0.92, 0.99)&-0.003&0.317&0.165 (0.173)&0.177 (-0.44, 0.62)&0.001&0.173\\\cline{2-10}
&\begin{tabular}{c}{\bf LIMLE: Full model}\end{tabular}&0.169 (0.231)&0.163 (-0.45, 0.76)&0.005&0.231&0.164 (0.139)&0.170 (-0.29, 0.63)&0.000&0.139\\\hline
\end{tabular}
}
\label{tab322}
\end{table}
\end{landscape}

\newpage
\begin{landscape}
\begin{table}[h]
\begin{center}
\caption{Summary of the results of model selection for each estimator (dichotomous and heterogeneous, normal dist.)}
\scalebox{0.8}{
\begin{tabular}{|c|c||c|c||c|c|}\hline
{{\bf Situation}}&{{\bf Method}}&\multicolumn{2}{|c||}{{\bf Sample size}: n=100}&\multicolumn{2}{|c|}{{\bf Sample size}: n=300}\\\cline{3-6}
&&
\begin{tabular}{c}
True model\\
n (\%)
\end{tabular}&
\begin{tabular}{c}
Including true model\\
n (\%)
\end{tabular}&
\begin{tabular}{c}
True model\\
n (\%)
\end{tabular}&
\begin{tabular}{c}
Including true model\\
n (\%)
\end{tabular}\\\hline\hline
\begin{tabular}{c}(1) Weak\\correlation,\\Strong\end{tabular}&\begin{tabular}{c}{\bf 2SRI}\end{tabular}&619 (61.9)&834 (83.4)&727 (72.7)&990 (99.0)\\\cline{2-6}
\begin{tabular}{c}IV, and\\Strong treatment\end{tabular}&\begin{tabular}{c}{\bf LIMLE: LAIC}\end{tabular}&594 (59.4)&825 (82.5)&764 (76.4)&988 (98.8)\\\cline{2-6}
&\begin{tabular}{c}{\bf LIMLE: LBIC}\end{tabular}&583 (58.3)&610 (61.0)&933 (93.3)&956 (95.6)\\\hline\hline
\begin{tabular}{c}(2) Weak\\correlation,\\Weak\end{tabular}&\begin{tabular}{c}{\bf 2SRI}\end{tabular}&619 (61.9)&828 (82.8)&699 (69.9)&993 (99.3)\\\cline{2-6}
\begin{tabular}{c}IV, and\\Strong treatment\end{tabular}&\begin{tabular}{c}{\bf LIMLE: LAIC}\end{tabular}&592 (59.2)&818 (81.8)&740 (74.0)&992 (99.2)\\\cline{2-6}
&\begin{tabular}{c}{\bf LIMLE: LBIC}\end{tabular}&581 (58.1)&602 (60.2)&933 (93.3)&958 (95.8)\\\hline\hline
\begin{tabular}{c}(3) Strong\\correlation,\\Strong\end{tabular}&\begin{tabular}{c}{\bf 2SRI}\end{tabular}&556 (55.6)&772 (77.2)&663 (66.3)&990 (99.0)\\\cline{2-6}
\begin{tabular}{c}IV, and\\Strong treatment\end{tabular}&\begin{tabular}{c}{\bf LIMLE: LAIC}\end{tabular}&578 (57.8)&764 (76.4)&772 (77.2)&988 (98.8)\\\cline{2-6}
&\begin{tabular}{c}{\bf LIMLE: LBIC}\end{tabular}&548 (54.8)&570 (57.0)&922 (92.2)&945 (94.5)\\\hline\hline
\begin{tabular}{c}(4) Weak\\correlation,\\Strong\end{tabular}&\begin{tabular}{c}{\bf 2SRI}\end{tabular}&641 (64.1)&843 (84.3)&716 (71.6)&995 (99.5)\\\cline{2-6}
\begin{tabular}{c}IV, and\\Weak treatment\end{tabular}&\begin{tabular}{c}{\bf LIMLE: LAIC}\end{tabular}&626 (62.6)&834 (83.4)&767 (76.7)&995 (99.5)\\\cline{2-6}
&\begin{tabular}{c}{\bf LIMLE: LBIC}\end{tabular}&616 (61.6)&641 (64.1)&942 (94.2)&973 (97.3)\\\hline
\end{tabular}
}
\label{tab323}
\end{center}
\end{table}
\end{landscape}

\subsubsection{Unspecified the correct joint distribution of unmeasured covariates}
Next, we consider that there are non-normal unmeasured covariates situation; specifically, there are two correlation structures of unmeasured covariates: t-copula and Clayton-copula. In this situation, it is hard to detect the correct joint distribution of unmeasured covariates accurately. In other words, both the 2SRI and the proposed methods, a bivariate normal distribution is assumed to estimate the causal effects and a probability of $Y_{1}$, nevertheless the true distribution has heavy-tailed joint distribution (t-copula) or asymmetric joint distribution (Clayton-copula). Therefore, the correct model is no longer included in the candidate models (i.e. misspecified the correlation structures). In the supplemental figures (Figure \ref{fig9} and Figure \ref{fig10}), each example of unmeasured covariates distribution is described both heavy-tailed joint distribution and asymmetric joint distribution situation. Under this setting, we confirm that our proposed model selection procedures and an ordinary 2SRI estimator with model selection work well or not, nevertheless there are more practical situation than the previous simulation setting.

\begin{landscape}
\begin{figure}[h]
\begin{center}
\begin{tabular}{c}
\includegraphics[width=24cm]{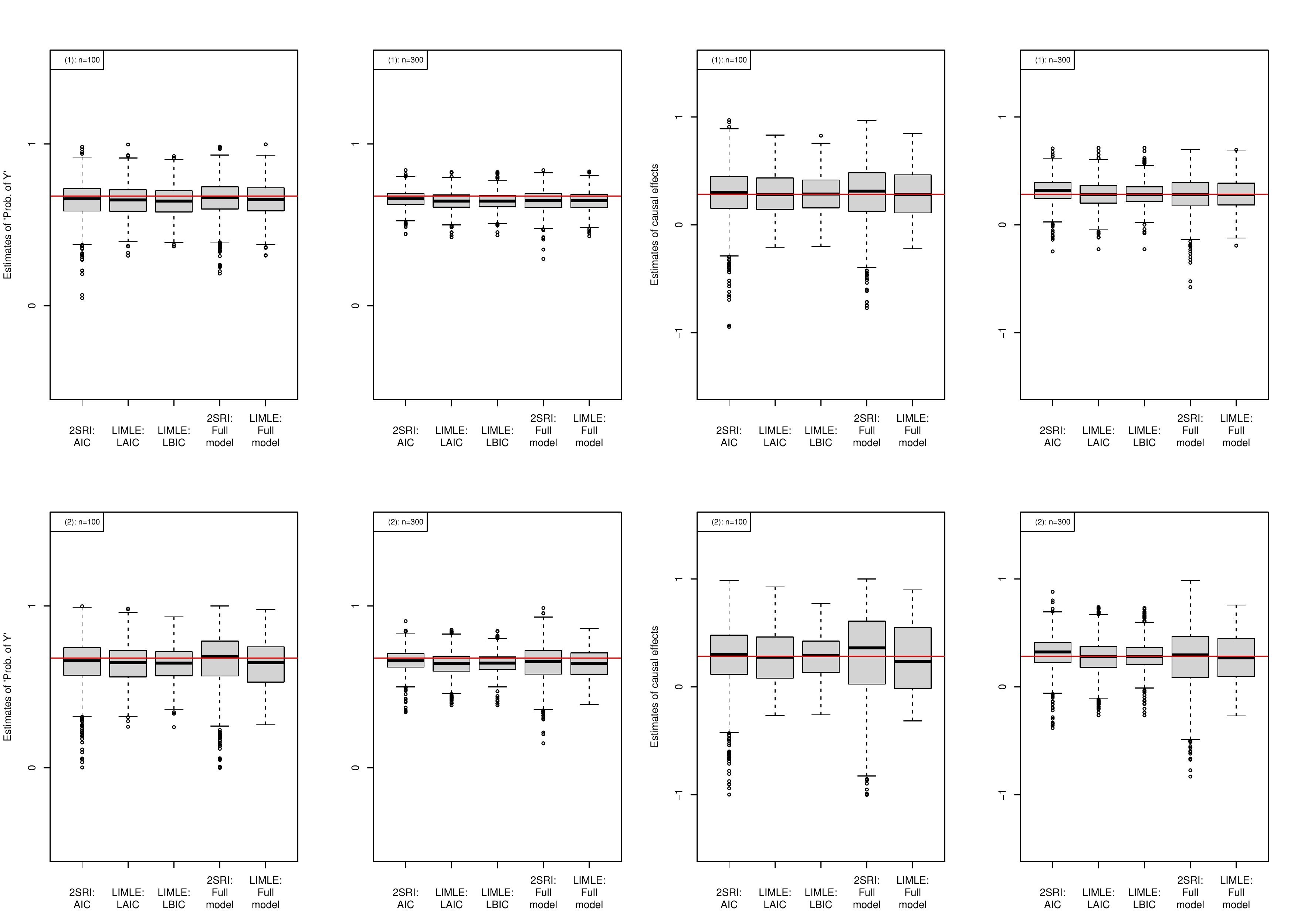}
\end{tabular}
\caption{Boxplots of descriptive statistics for each estimator (dichotomous and heterogeneous, logistic reg. \& t-copula) 1/2}
\label{fig5}
\end{center}
\end{figure}
\end{landscape}

\begin{landscape}
\begin{figure}[h]
\begin{center}
\begin{tabular}{c}
\includegraphics[width=24cm]{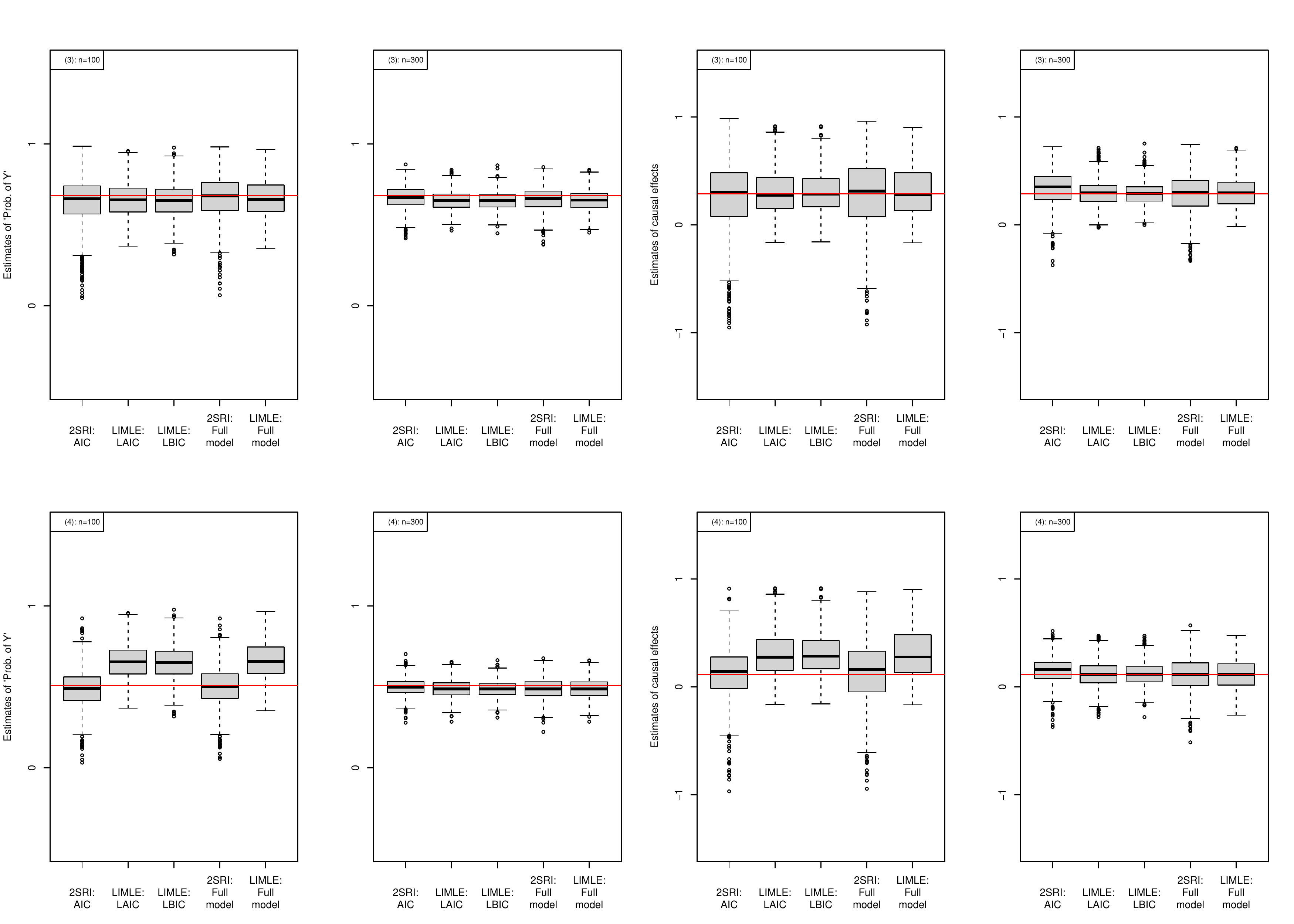}
\end{tabular}
\caption{Boxplots of descriptive statistics for each estimator (dichotomous and heterogeneous, logistic reg. \& t-copula) 2/2}
\label{fig6}
\end{center}
\end{figure}
\end{landscape}

At first, we show the results of heavy-tailed joint distribution situation. Results of the simulations are summarized in tables and supplemental figures. The estimated probability of $Y_{1}$ and causal effects are summarized in Table \ref{tab331}, Table \ref{tab332}, Figure \ref{fig5}, and Figure \ref{fig6}. Except for (4), the proposed procedure works well in terms of the accuracy and efficiency of these estimates. In particular, the result of LBIC is better in the sense of RMSE; whereas, the result of LAIC is somewhat high accuracy than LBIC. As mentioned previously, the 2SRI does not assume a dichotomous treatment situation, however, the 2SRI shows similar results as the proposed procedures; in some situations, the 2SRI is better. The results may come from the characteristic that the LIMLE is the full-likelihood based approach. Therefore, the LIMLE may tend to be affected by the model misspecification. On the other hand, the LBIC is the most efficient and has the most unbiased result in terms of the causal effects' estimates. In (4), our proposed method have large bias in small sample situation. In the large sample situation, the bias is improved; however, small sample's bias cannot be ignored. This may be derived from the unmeasured covariates distribution (affected by heavy-tailed distribution). Since there are only small treatment effects, the ``outlier" of unmeasured covariates (c.f. Figure \ref{fig9}) may distort the estimates more than in the case of large treatment effects.

The results of model selection are summarized in Table \ref{tab333}. the 2SRI and the LAIC can select the true model, or the models including the true model; however, the LBIC tends to select the underspecified models compared with the previous simulations. This is because the "true model" is no longer included in the candidate models in the sense that the correlation structure is misspecified. This is the same feature of the ordinary BIC (e.g. see Claeskens and Hjort, 2003).

\newpage
\begin{landscape}
\begin{table}[h]
\caption{Summary of causal effects for each estimator (prob. of $Y_{1}$, dichotomous and heterogeneous, logistic reg. \& t-copula.)}
\scalebox{0.8}{
\begin{tabular}{|c|c||c|c|c|c||c|c|c|c|}\hline
{\bf Situation}&{\bf Method}&\multicolumn{4}{|c||}{{\bf Sample size}: n=100}&\multicolumn{4}{|c|}{{\bf Sample size}: n=300}\\\cline{3-10}
&&{Mean (SD)}&{Median (Range)}&Bias&RMSE&{Mean (SD)}&{Median (Range)}&Bias&RMSE\\\hline\hline
\begin{tabular}{c}(1) Weak\\correlation,\end{tabular}&{{\bf 2SRI}}&0.653 (0.114)&0.661 (0.05, 0.98)&-0.024&0.116&0.660 (0.055)&0.660 (0.44, 0.84)&-0.018&0.058\\\cline{2-10}
\begin{tabular}{c}Strong\\IV, and\end{tabular}&\begin{tabular}{c}{\bf LIMLE: LAIC}\end{tabular}&0.650 (0.099)&0.654 (0.31, 1.00)&-0.027&0.103&0.647 (0.058)&0.647 (0.42, 0.83)&-0.031&0.065\\\cline{2-10}
\begin{tabular}{c}Strong\\treatment\end{tabular}&\begin{tabular}{c}{\bf LIMLE: LBIC}\end{tabular}&0.647 (0.095)&0.647 (0.37, 0.93)&-0.030&0.100&0.647 (0.053)&0.647 (0.44, 0.83)&-0.031&0.061\\\cline{2-10}
&\begin{tabular}{c}{\bf 2SRI: Full model}\end{tabular}&0.661 (0.118)&0.671 (0.20, 0.98)&-0.016&0.120&0.649 (0.067)&0.651 (0.29, 0.84)&-0.028&0.073\\\cline{2-10}
&\begin{tabular}{c}{\bf LIMLE: Full model}\end{tabular}&0.653 (0.105)&0.657 (0.31, 1.00)&-0.025&0.108&0.647 (0.063)&0.650 (0.43, 0.83)&-0.030&0.070\\\hline\hline

\begin{tabular}{c}(2) Weak\\correlation,\end{tabular}&{{\bf 2SRI}}&0.650 (0.143)&0.660 (0.00, 1.00)&-0.027&0.145&0.662 (0.069)&0.662 (0.34, 0.91)&-0.015&0.071\\\cline{2-10}
\begin{tabular}{c}Weak\\IV, and\end{tabular}&\begin{tabular}{c}{\bf LIMLE: LAIC}\end{tabular}&0.643 (0.117)&0.650 (0.25, 0.98)&-0.035&0.122&0.644 (0.077)&0.646 (0.39, 0.85)&-0.033&0.084\\\cline{2-10}
\begin{tabular}{c}Strong\\treatment\end{tabular}&\begin{tabular}{c}{\bf LIMLE: LBIC}\end{tabular}&0.641 (0.107)&0.647 (0.25, 0.93)&-0.036&0.113&0.647 (0.061)&0.648 (0.39, 0.84)&-0.031&0.068\\\cline{2-10}
&\begin{tabular}{c}{\bf 2SRI: Full model}\end{tabular}&0.661 (0.182)&0.687 (0.00, 1.00)&-0.016&0.183&0.648 (0.117)&0.657 (0.15, 0.99)&-0.029&0.121\\\cline{2-10}
&\begin{tabular}{c}{\bf LIMLE: Full model}\end{tabular}&0.637 (0.133)&0.650 (0.27, 0.98)&-0.041&0.139&0.641 (0.092)&0.645 (0.39, 0.86)&-0.036&0.099\\\hline\hline

\begin{tabular}{c}(3) Strong\\correlation,\end{tabular}&{{\bf 2SRI}}&0.645 (0.149)&0.662 (0.05, 0.99)&-0.035&0.153&0.668 (0.070)&0.670 (0.42, 0.87)&-0.012&0.071\\\cline{2-10}
\begin{tabular}{c}Strong\\IV, and\end{tabular}&\begin{tabular}{c}{\bf LIMLE: LAIC}\end{tabular}&0.657 (0.106)&0.655 (0.37, 0.96)&-0.023&0.109&0.651 (0.060)&0.651 (0.46, 0.84)&-0.029&0.066\\\cline{2-10}
\begin{tabular}{c}Strong\\treatment\end{tabular}&\begin{tabular}{c}{\bf LIMLE: LBIC}\end{tabular}&0.650 (0.104)&0.651 (0.32, 0.98)&-0.030&0.109&0.648 (0.055)&0.649 (0.45, 0.87)&-0.032&0.064\\\cline{2-10}
&\begin{tabular}{c}{\bf 2SRI: Full model}\end{tabular}&0.669 (0.136)&0.680 (0.07, 0.98)&-0.011&0.137&0.658 (0.072)&0.663 (0.38, 0.86)&-0.022&0.075\\\cline{2-10}
&\begin{tabular}{c}{\bf LIMLE: Full model}\end{tabular}&0.665 (0.111)&0.657 (0.35, 0.96)&-0.015&0.112&0.652 (0.065)&0.653 (0.45, 0.84)&-0.028&0.071\\\hline\hline

\begin{tabular}{c}(4) Weak\\correlation,\end{tabular}&{{\bf 2SRI}}&0.488 (0.114)&0.490 (0.03, 0.92)&-0.021&0.116&0.500 (0.055)&0.500 (0.28, 0.70)&-0.009&0.056\\\cline{2-10}
\begin{tabular}{c}Strong\\IV, and\end{tabular}&\begin{tabular}{c}{\bf LIMLE: LAIC}\end{tabular}&0.657 (0.106)&0.655 (0.37, 0.96)&0.148&0.182&0.488 (0.056)&0.488 (0.28, 0.65)&-0.021&0.059\\\cline{2-10}
\begin{tabular}{c}Weak\\treatment\end{tabular}&\begin{tabular}{c}{\bf LIMLE: LBIC}\end{tabular}&0.650 (0.104)&0.651 (0.32, 0.98)&0.141&0.175&0.487 (0.050)&0.487 (0.31, 0.66)&-0.022&0.055\\\cline{2-10}
&\begin{tabular}{c}{\bf 2SRI: Full model}\end{tabular}&0.500 (0.123)&0.504 (0.06, 0.92)&-0.009&0.123&0.488 (0.066)&0.488 (0.22, 0.68)&-0.021&0.069\\\cline{2-10}
&\begin{tabular}{c}{\bf LIMLE: Full model}\end{tabular}&0.665 (0.111)&0.657 (0.35, 0.96)&0.156&0.191&0.489 (0.060)&0.487 (0.28, 0.66)&-0.020&0.064\\\hline
\end{tabular}
}
\label{tab331}
\end{table}
\end{landscape}

\newpage
\begin{landscape}
\begin{table}[h]
\caption{Summary of causal effects for each estimator (causal effect, dichotomous and heterogeneous, logistic reg. \& t-copula.)}
\scalebox{0.8}{
\begin{tabular}{|c|c||c|c|c|c||c|c|c|c|}\hline
{\bf Situation}&{\bf Method}&\multicolumn{4}{|c||}{{\bf Sample size}: n=100}&\multicolumn{4}{|c|}{{\bf Sample size}: n=300}\\\cline{3-10}
&&{Mean (SD)}&{Median (Range)}&Bias&RMSE&{Mean (SD)}&{Median (Range)}&Bias&RMSE\\\hline\hline
\begin{tabular}{c}(1) Weak\\correlation,\end{tabular}&{{\bf 2SRI}}&0.285 (0.245)&0.303 (-0.95, 0.97)&0.001&0.245&0.316 (0.123)&0.320 (-0.25, 0.71)&0.032&0.127\\\cline{2-10}
\begin{tabular}{c}Strong\\IV, and\end{tabular}&\begin{tabular}{c}{\bf LIMLE: LAIC}\end{tabular}&0.288 (0.206)&0.279 (-0.21, 0.83)&0.004&0.207&0.284 (0.129)&0.280 (-0.22, 0.71)&0.000&0.129\\\cline{2-10}
\begin{tabular}{c}Strong\\treatment\end{tabular}&\begin{tabular}{c}{\bf LIMLE: LBIC}\end{tabular}&0.290 (0.182)&0.287 (-0.20, 0.83)&0.006&0.182&0.285 (0.109)&0.282 (-0.22, 0.71)&0.001&0.109\\\cline{2-10}
&\begin{tabular}{c}{\bf 2SRI: Full model}\end{tabular}&0.289 (0.286)&0.315 (-0.77, 0.97)&0.005&0.286&0.277 (0.168)&0.278 (-0.58, 0.70)&-0.007&0.168\\\cline{2-10}
&\begin{tabular}{c}{\bf LIMLE: Full model}\end{tabular}&0.286 (0.232)&0.283 (-0.22, 0.85)&0.002&0.233&0.282 (0.146)&0.277 (-0.19, 0.70)&-0.002&0.146\\\hline\hline

\begin{tabular}{c}(2) Weak\\correlation,\end{tabular}&{{\bf 2SRI}}&0.275 (0.316)&0.298 (-1.00, 0.99)&-0.009&0.316&0.319 (0.165)&0.323 (-0.38, 0.88)&0.035&0.168\\\cline{2-10}
\begin{tabular}{c}Weak\\IV, and\end{tabular}&\begin{tabular}{c}{\bf LIMLE: LAIC}\end{tabular}&0.274 (0.249)&0.276 (-0.26, 0.93)&-0.010&0.249&0.279 (0.178)&0.281 (-0.26, 0.74)&-0.005&0.178\\\cline{2-10}
\begin{tabular}{c}Strong\\treatment\end{tabular}&\begin{tabular}{c}{\bf LIMLE: LBIC}\end{tabular}&0.281 (0.206)&0.289 (-0.26, 0.77)&-0.003&0.206&0.285 (0.130)&0.284 (-0.26, 0.73)&0.001&0.130\\\cline{2-10}
&\begin{tabular}{c}{\bf 2SRI: Full model}\end{tabular}&0.293 (0.431)&0.361 (-1.00, 1.00)&0.009&0.431&0.270 (0.301)&0.297 (-0.83, 0.98)&-0.014&0.301\\\cline{2-10}
&\begin{tabular}{c}{\bf LIMLE: Full model}\end{tabular}&0.257 (0.304)&0.238 (-0.31, 0.90)&-0.027&0.305&0.270 (0.224)&0.270 (-0.27, 0.76)&-0.014&0.225\\\hline\hline

\begin{tabular}{c}(3) Strong\\correlation,\end{tabular}&{{\bf 2SRI}}&0.250 (0.331)&0.298 (-0.95, 0.98)&-0.037&0.333&0.334 (0.165)&0.353 (-0.37, 0.72)&0.046&0.171\\\cline{2-10}
\begin{tabular}{c}Strong\\IV, and\end{tabular}&\begin{tabular}{c}{\bf LIMLE: LAIC}\end{tabular}&0.299 (0.205)&0.275 (-0.17, 0.91)&0.012&0.205&0.296 (0.123)&0.295 (-0.03, 0.71)&0.009&0.124\\\cline{2-10}
\begin{tabular}{c}Strong\\treatment\end{tabular}&\begin{tabular}{c}{\bf LIMLE: LBIC}\end{tabular}&0.300 (0.184)&0.284 (-0.16, 0.91)&0.012&0.184&0.292 (0.105)&0.290 (0.00, 0.75)&0.004&0.105\\\cline{2-10}
&\begin{tabular}{c}{\bf 2SRI: Full model}\end{tabular}&0.280 (0.330)&0.312 (-0.92, 0.96)&-0.008&0.330&0.286 (0.185)&0.303 (-0.33, 0.75)&-0.001&0.185\\\cline{2-10}
&\begin{tabular}{c}{\bf LIMLE: Full model}\end{tabular}&0.312 (0.229)&0.277 (-0.17, 0.90)&0.024&0.231&0.297 (0.142)&0.297 (-0.01, 0.71)&0.010&0.142\\\hline\hline

\begin{tabular}{c}(4) Weak\\correlation,\end{tabular}&{{\bf 2SRI}}&0.121 (0.241)&0.142 (-0.97, 0.91)&0.005&0.241&0.153 (0.120)&0.160 (-0.37, 0.52)&0.037&0.125\\\cline{2-10}
\begin{tabular}{c}Strong\\IV, and\end{tabular}&\begin{tabular}{c}{\bf LIMLE: LAIC}\end{tabular}&0.299 (0.205)&0.275 (-0.17, 0.91)&0.183&0.275&0.119 (0.126)&0.114 (-0.28, 0.47)&0.004&0.126\\\cline{2-10}
\begin{tabular}{c}Weak\\treatment\end{tabular}&\begin{tabular}{c}{\bf LIMLE: LBIC}\end{tabular}&0.300 (0.184)&0.284 (-0.16, 0.91)&0.184&0.260&0.120 (0.104)&0.117 (-0.28, 0.47)&0.004&0.104\\\cline{2-10}
&\begin{tabular}{c}{\bf 2SRI: Full model}\end{tabular}&0.136 (0.286)&0.162 (-0.94, 0.88)&0.020&0.286&0.112 (0.164)&0.116 (-0.51, 0.57)&-0.003&0.164\\\cline{2-10}
&\begin{tabular}{c}{\bf LIMLE: Full model}\end{tabular}&0.312 (0.229)&0.277 (-0.17, 0.90)&0.196&0.302&0.118 (0.142)&0.115 (-0.26, 0.48)&0.002&0.142\\\hline
\end{tabular}
}
\label{tab332}
\end{table}
\end{landscape}

\newpage
\begin{landscape}
\begin{table}[h]
\begin{center}
\caption{Summary of the results of model selection for each estimator (dichotomous and heterogeneous, logistic reg. \& t-copula.)}
\scalebox{0.8}{
\begin{tabular}{|c|c||c|c||c|c|}\hline
{{\bf Situation}}&{{\bf Method}}&\multicolumn{2}{|c||}{{\bf Sample size}: n=100}&\multicolumn{2}{|c|}{{\bf Sample size}: n=300}\\\cline{3-6}
&&
\begin{tabular}{c}
True model\\
n (\%)
\end{tabular}&
\begin{tabular}{c}
Including true model\\
n (\%)
\end{tabular}&
\begin{tabular}{c}
True model\\
n (\%)
\end{tabular}&
\begin{tabular}{c}
Including true model\\
n (\%)
\end{tabular}\\\hline\hline
\begin{tabular}{c}(1) Weak\\correlation,\\Strong\end{tabular}&\begin{tabular}{c}{\bf 2SRI}\end{tabular}&465 (46.5)&625 (62.5)&726 (72.6)&947 (94.7)\\\cline{2-6}
\begin{tabular}{c}IV, and\\Strong treatment\end{tabular}&\begin{tabular}{c}{\bf LIMLE: LAIC}\end{tabular}&486 (48.6)&634 (63.4)&753 (75.3)&950 (95.0)\\\cline{2-6}
&\begin{tabular}{c}{\bf LIMLE: LBIC}\end{tabular}&355 (35.5)&368 (36.8)&757 (75.7)&774 (77.4)\\\hline\hline

\begin{tabular}{c}(2) Weak\\correlation,\\Weak\end{tabular}&\begin{tabular}{c}{\bf 2SRI}\end{tabular}&455 (45.5)&631 (63.1)&700 (70.0)&949 (94.9)\\\cline{2-6}
\begin{tabular}{c}IV, and\\Strong treatment\end{tabular}&\begin{tabular}{c}{\bf LIMLE: LAIC}\end{tabular}&480 (48.0)&630 (63.0)&735 (73.5)&947 (94.7)\\\cline{2-6}
&\begin{tabular}{c}{\bf LIMLE: LBIC}\end{tabular}&328 (32.8)&340 (34.0)&740 (74.0)&756 (75.6)\\\hline\hline

\begin{tabular}{c}(3) Strong\\correlation,\\Strong\end{tabular}&\begin{tabular}{c}{\bf 2SRI}\end{tabular}&473 (47.3)&636 (63.6)&648 (64.8)&925 (92.5)\\\cline{2-6}
\begin{tabular}{c}IV, and\\Strong treatment\end{tabular}&\begin{tabular}{c}{\bf LIMLE: LAIC}\end{tabular}&491 (49.1)&647 (64.7)&731 (73.1)&937 (93.7)\\\cline{2-6}
&\begin{tabular}{c}{\bf LIMLE: LBIC}\end{tabular}&351 (35.1)&364 (36.4)&740 (74.0)&752 (75.2)\\\hline\hline

\begin{tabular}{c}(4) Weak\\correlation,\\Strong\end{tabular}&\begin{tabular}{c}{\bf 2SRI}\end{tabular}&515 (51.5)&669 (66.9)&717 (71.7)&958 (95.8)\\\cline{2-6}
\begin{tabular}{c}IV, and\\Weak treatment\end{tabular}&\begin{tabular}{c}{\bf LIMLE: LAIC}\end{tabular}&491 (49.1)&647 (64.7)&740 (74.0)&961 (96.1)\\\cline{2-6}
&\begin{tabular}{c}{\bf LIMLE: LBIC}\end{tabular}&351 (35.1)&364 (36.4)&807 (80.7)&823 (82.3)\\\hline
\end{tabular}
}
\label{tab333}
\end{center}
\end{table}
\end{landscape}

\begin{landscape}
\begin{figure}[h]
\begin{center}
\begin{tabular}{c}
\includegraphics[width=24cm]{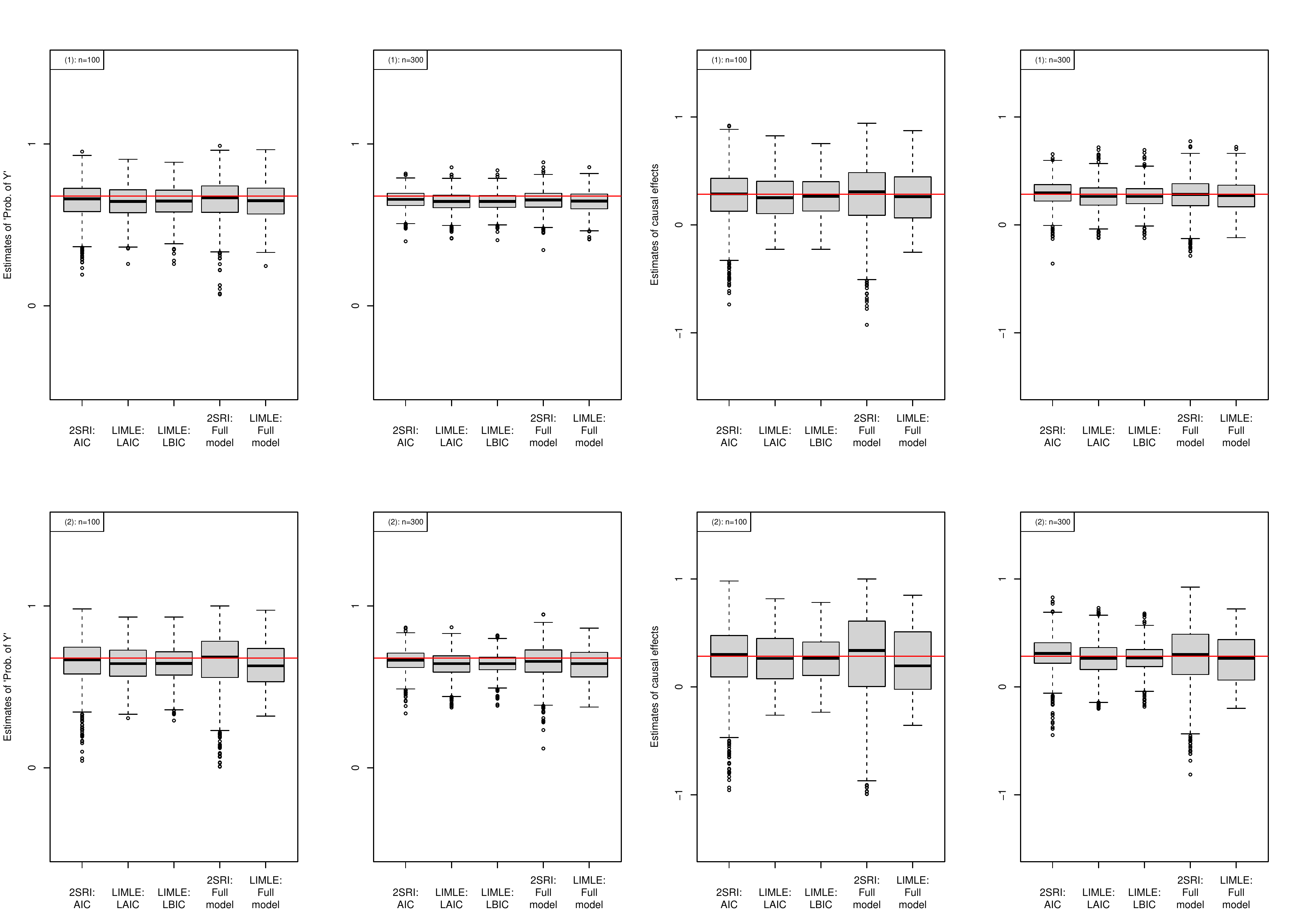}
\end{tabular}
\caption{Boxplots of descriptive statistics for each estimator (dichotomous and heterogeneous, logistic reg. \& Clayton-copula) 1/2}
\label{fig7}
\end{center}
\end{figure}
\end{landscape}

\begin{landscape}
\begin{figure}[h]
\begin{center}
\begin{tabular}{c}
\includegraphics[width=24cm]{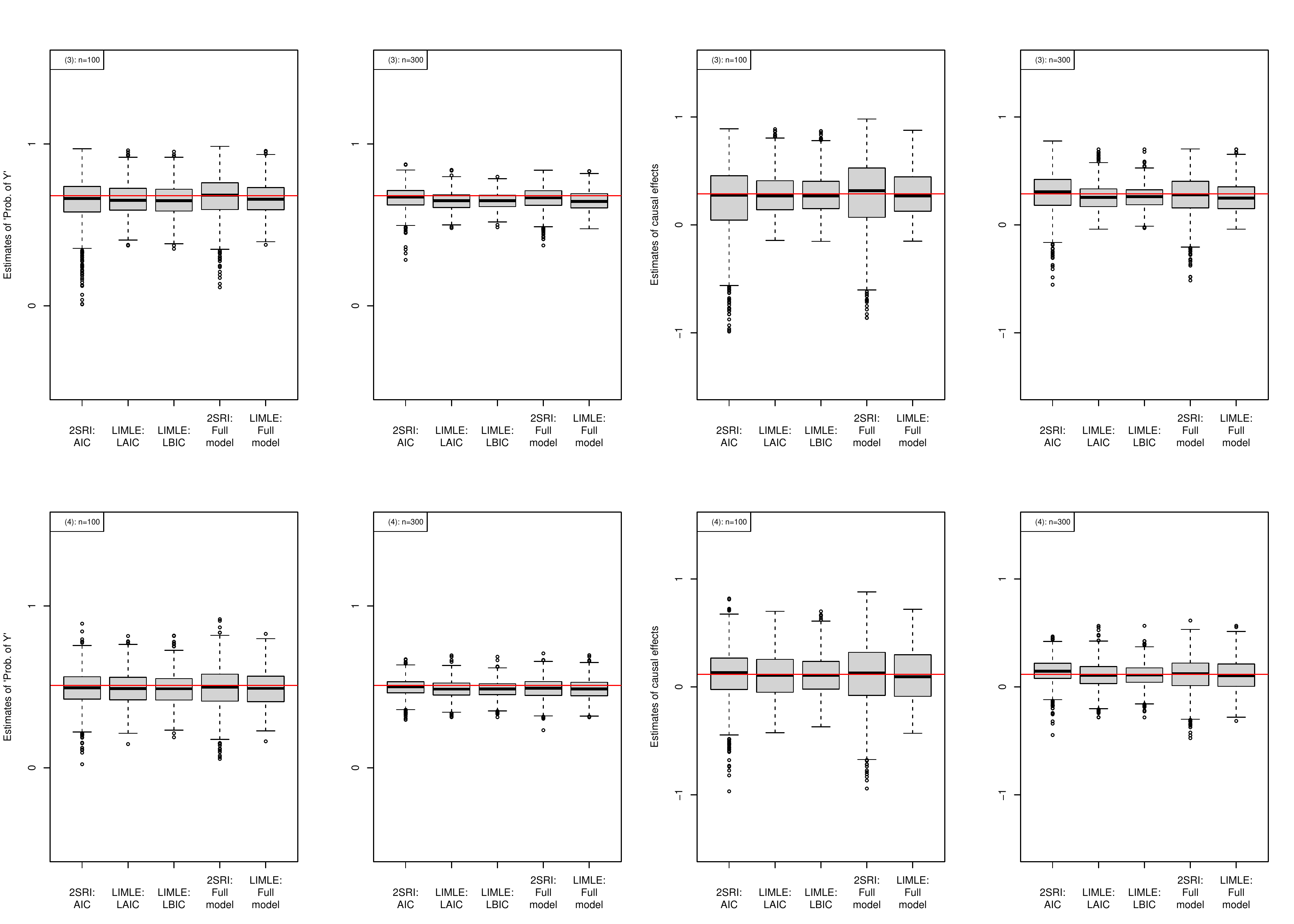}
\end{tabular}
\caption{Boxplots of descriptive statistics for each estimator (dichotomous and heterogeneous, logistic reg. \& Clayton-copula) 2/2}
\label{fig8}
\end{center}
\end{figure}
\end{landscape}

Next, we show the results of asymmetric joint distribution situation. Results of the simulations are summarized in tables and supplemental figures. The estimated probability of $Y_{1}$ and causal effects are summarized in Table \ref{tab341}, Table \ref{tab342}, Figure \ref{fig7}, and Figure \ref{fig8}. Also, results of the model selection is summarized in Table \ref{tab343}. Surprisingly, both the 2SRI and the LIMLE estimates have only small biases even when the bivariate normal distribution (symmetric distribution) is assumed for unmeasured covariates. Note that the situation in (4), which had a bias in the previous simulation, could be estimated without bias. Except for this interesting result, both the estimates and model selection, all simulation results are similar as the previous one: both LIML and 2SRI work well in the sense of the estimated probability of $Y_{1}$, and the LBIC is the best in the sense of the estimated causal effects. Also, the results of model selection tend to be better than heavy-tailed situation.

From the above, when we can specify the correct joint distribution of unmeasured covariates, not only using a LIMLE with model selection procedures but also a 2SRI with model selection procedures are good choice to estimate the marginal distribution of an outcome and the causal effects; especially, the latter result are different from Basu et al. (2017). Also, the simulation results show that the 2SRI works to some extent even if there are a dichotomous treatment situation where the 2SRI does not assume the situation; to the best of my knowledge, this is new findings. Even when we cannot specify the correct joint distribution of unmeasured covariates, both a LIMLE and a 2SRI show a robust results in the sense that the estimates does not have large bias or low efficiency; a LIMLE tends to be affected by the model misspecification since the method is full-likelihood approach. Also, a LIMLE may be affected by the ``outlier" of unmeasured covariates when there are only small causal effects. On the other hand, even when there are asymmetric unmeasured covariates, there are robust results; the results are hard to give some interpretation, however, this is the feature of these methods. We need to verify the situation, and confirm the feature the methods for the future works.

\newpage
\begin{landscape}
\begin{table}[h]
\caption{Summary of causal effects for each estimator (prob. of $Y_{1}$, dichotomous and heterogeneous, logistic reg. \& Clayton-copula.)}
\scalebox{0.8}{
\begin{tabular}{|c|c||c|c|c|c||c|c|c|c|}\hline
{\bf Situation}&{\bf Method}&\multicolumn{4}{|c||}{{\bf Sample size}: n=100}&\multicolumn{4}{|c|}{{\bf Sample size}: n=300}\\\cline{3-10}
&&{Mean (SD)}&{Median (Range)}&Bias&RMSE&{Mean (SD)}&{Median (Range)}&Bias&RMSE\\\hline\hline
\begin{tabular}{c}(1) Weak\\correlation,\end{tabular}&{{\bf 2SRI}}&0.649 (0.115)&0.660 (0.19, 0.95)&-0.029&0.118&0.655 (0.057)&0.658 (0.40, 0.82)&-0.023&0.061\\\cline{2-10}
\begin{tabular}{c}Strong\\IV, and\end{tabular}&\begin{tabular}{c}{\bf LIMLE: LAIC}\end{tabular}&0.644 (0.103)&0.646 (0.26, 0.91)&-0.033&0.109&0.642 (0.060)&0.645 (0.42, 0.86)&-0.035&0.069\\\cline{2-10}
\begin{tabular}{c}Strong\\treatment\end{tabular}&\begin{tabular}{c}{\bf LIMLE: LBIC}\end{tabular}&0.644 (0.098)&0.647 (0.26, 0.89)&-0.034&0.104&0.644 (0.055)&0.645 (0.41, 0.84)&-0.034&0.065\\\cline{2-10}
&\begin{tabular}{c}{\bf 2SRI: Full model}\end{tabular}&0.653 (0.143)&0.667 (0.04, 0.98)&-0.025&0.145&0.661 (0.070)&0.666 (0.34, 0.87)&-0.016&0.072\\\cline{2-10}
&\begin{tabular}{c}{\bf LIMLE: Full model}\end{tabular}&0.645 (0.111)&0.650 (0.25, 0.96)&-0.033&0.116&0.643 (0.066)&0.647 (0.41, 0.86)&-0.035&0.074\\\hline\hline

\begin{tabular}{c}(2) Weak\\correlation,\end{tabular}&{{\bf 2SRI}}&0.723 (0.115)&0.735 (0.16, 1.00)&0.008&0.115&0.732 (0.053)&0.732 (0.51, 0.88)&0.017&0.056\\\cline{2-10}
\begin{tabular}{c}Weak\\IV, and\end{tabular}&\begin{tabular}{c}{\bf LIMLE: LAIC}\end{tabular}&0.644 (0.115)&0.644 (0.31, 0.93)&-0.033&0.120&0.639 (0.082)&0.644 (0.37, 0.87)&-0.038&0.091\\\cline{2-10}
\begin{tabular}{c}Strong\\treatment\end{tabular}&\begin{tabular}{c}{\bf LIMLE: LBIC}\end{tabular}&0.642 (0.107)&0.645 (0.29, 0.93)&-0.035&0.112&0.643 (0.063)&0.644 (0.38, 0.82)&-0.034&0.071\\\cline{2-10}
&\begin{tabular}{c}{\bf 2SRI: Full model}\end{tabular}&0.660 (0.182)&0.684 (0.01, 1.00)&-0.018&0.183&0.652 (0.110)&0.658 (0.12, 0.95)&-0.026&0.113\\\cline{2-10}
&\begin{tabular}{c}{\bf LIMLE: Full model}\end{tabular}&0.633 (0.130)&0.630 (0.32, 0.97)&-0.044&0.138&0.635 (0.099)&0.644 (0.38, 0.86)&-0.042&0.108\\\hline\hline

\begin{tabular}{c}(3) Strong\\correlation,\end{tabular}&{{\bf 2SRI}}&0.645 (0.142)&0.662 (0.01, 0.97)&-0.035&0.146&0.665 (0.071)&0.671 (0.28, 0.87)&-0.015&0.073\\\cline{2-10}
\begin{tabular}{c}Strong\\IV, and\end{tabular}&\begin{tabular}{c}{\bf LIMLE: LAIC}\end{tabular}&0.660 (0.097)&0.652 (0.37, 0.96)&-0.020&0.099&0.649 (0.058)&0.649 (0.48, 0.84)&-0.031&0.065\\\cline{2-10}
\begin{tabular}{c}Strong\\treatment\end{tabular}&\begin{tabular}{c}{\bf LIMLE: LBIC}\end{tabular}&0.653 (0.099)&0.649 (0.35, 0.95)&-0.027&0.102&0.650 (0.053)&0.650 (0.49, 0.80)&-0.030&0.061\\\cline{2-10}
&\begin{tabular}{c}{\bf 2SRI: Full model}\end{tabular}&0.672 (0.130)&0.684 (0.11, 0.98)&-0.008&0.130&0.663 (0.069)&0.667 (0.37, 0.84)&-0.016&0.071\\\cline{2-10}
&\begin{tabular}{c}{\bf LIMLE: Full model}\end{tabular}&0.664 (0.099)&0.659 (0.38, 0.96)&-0.016&0.100&0.649 (0.062)&0.645 (0.48, 0.83)&-0.031&0.070\\\hline\hline

\begin{tabular}{c}(4) Weak\\correlation,\end{tabular}&{{\bf 2SRI}}&0.488 (0.112)&0.495 (0.02, 0.89)&-0.021&0.114&0.497 (0.056)&0.500 (0.30, 0.67)&-0.012&0.057\\\cline{2-10}
\begin{tabular}{c}Strong\\IV, and\end{tabular}&\begin{tabular}{c}{\bf LIMLE: LAIC}\end{tabular}&0.489 (0.103)&0.490 (0.15, 0.81)&-0.020&0.105&0.486 (0.058)&0.486 (0.31, 0.69)&-0.023&0.062\\\cline{2-10}
\begin{tabular}{c}Weak\\treatment\end{tabular}&\begin{tabular}{c}{\bf LIMLE: LBIC}\end{tabular}&0.485 (0.096)&0.489 (0.19, 0.82)&-0.024&0.099&0.486 (0.053)&0.487 (0.31, 0.69)&-0.023&0.058\\\cline{2-10}
&\begin{tabular}{c}{\bf 2SRI: Full model}\end{tabular}&0.495 (0.130)&0.499 (0.06, 0.92)&-0.014&0.131&0.490 (0.066)&0.492 (0.23, 0.71)&-0.019&0.069\\\cline{2-10}
&\begin{tabular}{c}{\bf LIMLE: Full model}\end{tabular}&0.491 (0.109)&0.491 (0.16, 0.83)&-0.018&0.111&0.487 (0.063)&0.488 (0.31, 0.70)&-0.022&0.067\\\hline
\end{tabular}
}
\label{tab341}
\end{table}
\end{landscape}

\newpage
\begin{landscape}
\begin{table}[h]
\caption{Summary of causal effects for each estimator (causal effect, dichotomous and heterogeneous, logistic reg. \& Clayton-copula.)}
\scalebox{0.8}{
\begin{tabular}{|c|c||c|c|c|c||c|c|c|c|}\hline
{\bf Situation}&{\bf Method}&\multicolumn{4}{|c||}{{\bf Sample size}: n=100}&\multicolumn{4}{|c|}{{\bf Sample size}: n=300}\\\cline{3-10}
&&{Mean (SD)}&{Median (Range)}&Bias&RMSE&{Mean (SD)}&{Median (Range)}&Bias&RMSE\\\hline\hline
\begin{tabular}{c}(1) Weak\\correlation,\end{tabular}&{{\bf 2SRI}}&0.257 (0.251)&0.287 (-0.74, 0.92)&-0.027&0.253&0.293 (0.121)&0.296 (-0.36, 0.66)&0.009&0.122\\\cline{2-10}
\begin{tabular}{c}Strong\\IV, and\end{tabular}&\begin{tabular}{c}{\bf LIMLE: LAIC}\end{tabular}&0.259 (0.208)&0.252 (-0.23, 0.83)&-0.025&0.209&0.263 (0.127)&0.264 (-0.12, 0.72)&-0.021&0.128\\\cline{2-10}
\begin{tabular}{c}Strong\\treatment\end{tabular}&\begin{tabular}{c}{\bf LIMLE: LBIC}\end{tabular}&0.266 (0.183)&0.265 (-0.23, 0.75)&-0.018&0.184&0.266 (0.108)&0.264 (-0.12, 0.69)&-0.018&0.110\\\cline{2-10}
&\begin{tabular}{c}{\bf 2SRI: Full model}\end{tabular}&0.268 (0.306)&0.308 (-0.93, 0.94)&-0.016&0.306&0.274 (0.160)&0.282 (-0.29, 0.77)&-0.010&0.160\\\cline{2-10}
&\begin{tabular}{c}{\bf LIMLE: Full model}\end{tabular}&0.261 (0.239)&0.262 (-0.25, 0.87)&-0.023&0.240&0.265 (0.145)&0.271 (-0.12, 0.72)&-0.019&0.146\\\hline\hline

\begin{tabular}{c}(2) Weak\\correlation,\end{tabular}&{{\bf 2SRI}}&0.265 (0.323)&0.301 (-0.96, 0.98)&-0.019&0.323&0.306 (0.165)&0.309 (-0.45, 0.83)&0.022&0.166\\\cline{2-10}
\begin{tabular}{c}Weak\\IV, and\end{tabular}&\begin{tabular}{c}{\bf LIMLE: LAIC}\end{tabular}&0.264 (0.240)&0.264 (-0.26, 0.82)&-0.020&0.241&0.259 (0.185)&0.265 (-0.20, 0.73)&-0.025&0.187\\\cline{2-10}
\begin{tabular}{c}Strong\\treatment\end{tabular}&\begin{tabular}{c}{\bf LIMLE: LBIC}\end{tabular}&0.265 (0.205)&0.265 (-0.24, 0.78)&-0.019&0.206&0.266 (0.125)&0.269 (-0.18, 0.68)&-0.018&0.127\\\cline{2-10}
&\begin{tabular}{c}{\bf 2SRI: Full model}\end{tabular}&0.278 (0.436)&0.336 (-0.99, 1.00)&-0.006&0.436&0.276 (0.288)&0.300 (-0.81, 0.92)&-0.008&0.288\\\cline{2-10}
&\begin{tabular}{c}{\bf LIMLE: Full model}\end{tabular}&0.239 (0.291)&0.195 (-0.36, 0.85)&-0.045&0.295&0.255 (0.234)&0.264 (-0.20, 0.72)&-0.029&0.236\\\hline\hline

\begin{tabular}{c}(3) Strong\\correlation,\end{tabular}&{{\bf 2SRI}}&0.225 (0.332)&0.275 (-0.99, 0.89)&-0.063&0.338&0.289 (0.189)&0.306 (-0.55, 0.78)&0.001&0.189\\\cline{2-10}
\begin{tabular}{c}Strong\\IV, and\end{tabular}&\begin{tabular}{c}{\bf LIMLE: LAIC}\end{tabular}&0.289 (0.196)&0.268 (-0.14, 0.89)&0.002&0.196&0.259 (0.126)&0.256 (-0.04, 0.70)&-0.028&0.129\\\cline{2-10}
\begin{tabular}{c}Strong\\treatment\end{tabular}&\begin{tabular}{c}{\bf LIMLE: LBIC}\end{tabular}&0.288 (0.179)&0.269 (-0.15, 0.87)&0.001&0.179&0.262 (0.105)&0.261 (-0.03, 0.70)&-0.026&0.108\\\cline{2-10}
&\begin{tabular}{c}{\bf 2SRI: Full model}\end{tabular}&0.273 (0.343)&0.317 (-0.86, 0.98)&-0.014&0.343&0.264 (0.191)&0.278 (-0.51, 0.70)&-0.024&0.193\\\cline{2-10}
&\begin{tabular}{c}{\bf LIMLE: Full model}\end{tabular}&0.300 (0.215)&0.268 (-0.15, 0.88)&0.013&0.216&0.261 (0.144)&0.247 (-0.04, 0.70)&-0.026&0.146\\\hline\hline

\begin{tabular}{c}(4) Weak\\correlation,\end{tabular}&{{\bf 2SRI}}&0.107 (0.244)&0.133 (-0.97, 0.82)&-0.009&0.244&0.141 (0.118)&0.147 (-0.45, 0.47)&0.026&0.121\\\cline{2-10}
\begin{tabular}{c}Strong\\IV, and\end{tabular}&\begin{tabular}{c}{\bf LIMLE: LAIC}\end{tabular}&0.110 (0.212)&0.106 (-0.42, 0.70)&-0.005&0.212&0.108 (0.126)&0.109 (-0.28, 0.57)&-0.008&0.126\\\cline{2-10}
\begin{tabular}{c}Weak\\treatment\end{tabular}&\begin{tabular}{c}{\bf LIMLE: LBIC}\end{tabular}&0.114 (0.184)&0.109 (-0.37, 0.70)&-0.002&0.184&0.110 (0.106)&0.110 (-0.28, 0.57)&-0.006&0.106\\\cline{2-10}
&\begin{tabular}{c}{\bf 2SRI: Full model}\end{tabular}&0.110 (0.305)&0.127 (-0.94, 0.88)&-0.006&0.305&0.115 (0.162)&0.123 (-0.47, 0.61)&-0.001&0.162\\\cline{2-10}
&\begin{tabular}{c}{\bf LIMLE: Full model}\end{tabular}&0.109 (0.240)&0.095 (-0.43, 0.72)&-0.007&0.240&0.109 (0.145)&0.104 (-0.32, 0.57)&-0.007&0.146\\\hline
\end{tabular}
}
\label{tab342}
\end{table}
\end{landscape}

\newpage
\begin{landscape}
\begin{table}[h]
\begin{center}
\caption{Summary of the results of model selection for each estimator (dichotomous and heterogeneous, logistic reg. \& Clayton-copula.)}
\scalebox{0.8}{
\begin{tabular}{|c|c||c|c||c|c|}\hline
{{\bf Situation}}&{{\bf Method}}&\multicolumn{2}{|c||}{{\bf Sample size}: n=100}&\multicolumn{2}{|c|}{{\bf Sample size}: n=300}\\\cline{3-6}
&&
\begin{tabular}{c}
True model\\
n (\%)
\end{tabular}&
\begin{tabular}{c}
Including true model\\
n (\%)
\end{tabular}&
\begin{tabular}{c}
True model\\
n (\%)
\end{tabular}&
\begin{tabular}{c}
Including true model\\
n (\%)
\end{tabular}\\\hline\hline
\begin{tabular}{c}(1) Weak\\correlation,\\Strong\end{tabular}&\begin{tabular}{c}{\bf 2SRI}\end{tabular}&541 (54.1)&717 (71.7)&749 (74.9)&984 (98.4)\\\cline{2-6}
\begin{tabular}{c}IV, and\\Strong treatment\end{tabular}&\begin{tabular}{c}{\bf LIMLE: LAIC}\end{tabular}&565 (56.5)&737 (73.7)&783 (78.3)&984 (98.4)\\\cline{2-6}
&\begin{tabular}{c}{\bf LIMLE: LBIC}\end{tabular}&423 (42.3)&438 (43.8)&861 (86.1)&877 (87.7)\\\hline\hline

\begin{tabular}{c}(2) Weak\\correlation,\\Weak\end{tabular}&\begin{tabular}{c}{\bf 2SRI}\end{tabular}&527 (52.7)&719 (71.9)&724 (72.4)&984 (98.4)\\\cline{2-6}
\begin{tabular}{c}IV, and\\Strong treatment\end{tabular}&\begin{tabular}{c}{\bf LIMLE: LAIC}\end{tabular}&564 (56.4)&731 (73.1)&753 (75.3)&984 (98.4)\\\cline{2-6}
&\begin{tabular}{c}{\bf LIMLE: LBIC}\end{tabular}&444 (44.4)&456 (45.6)&849 (84.9)&866 (86.6)\\\hline\hline

\begin{tabular}{c}(3) Strong\\correlation,\\Strong\end{tabular}&\begin{tabular}{c}{\bf 2SRI}\end{tabular}&531 (53.1)&759 (75.9)&670 (67.0)&989 (98.9)\\\cline{2-6}
\begin{tabular}{c}IV, and\\Strong treatment\end{tabular}&\begin{tabular}{c}{\bf LIMLE: LAIC}\end{tabular}&589 (58.9)&773 (77.3)&757 (75.7)&992 (99.2)\\\cline{2-6}
&\begin{tabular}{c}{\bf LIMLE: LBIC}\end{tabular}&478 (47.8)&500 (50.0)&921 (92.1)&938 (93.8)\\\hline\hline

\begin{tabular}{c}(4) Weak\\correlation,\\Strong\end{tabular}&\begin{tabular}{c}{\bf 2SRI}\end{tabular}&548 (54.8)&733 (73.3)&763 (76.3)&990 (99.0)\\\cline{2-6}
\begin{tabular}{c}IV, and\\Weak treatment\end{tabular}&\begin{tabular}{c}{\bf LIMLE: LAIC}\end{tabular}&557 (55.7)&743 (74.3)&794 (79.4)&992 (99.2)\\\cline{2-6}
&\begin{tabular}{c}{\bf LIMLE: LBIC}\end{tabular}&461 (46.1)&484 (48.4)&888 (88.8)&907 (90.7)\\\hline
\end{tabular}
}
\label{tab343}
\end{center}
\end{table}
\end{landscape}
\section{Conclusions and Future Work}\noindent
In this article, a binary outcome model with unmeasured covariates is considered. Two-stage residual inclusion (2SRI) is applied to this situation, however, it may be derive some biased estimates (Basu et al., 2017). Therefore, limited-information maximum likelihood (LIML), which has similar features to 2SRI, is the main consideration in this article. Since LIML is full-likelihood approach, a model selection procedure like the AIC or BIC can be considered naturally; we propose LIML-based model selection procedures referred to as the LAIC and LBIC. Also, we show that the properties of the proposed procedures not only numerically but also theoretically; the procedures have the same features as AIC and BIC, respectively. In particular, we prove model selection consistency of the later BIC-type model selection criterion. From the simulation results, both the LAIC and the LBIC work well when an unmeasured covariate distribution is specified correctly. In addition, when we cannot specify the correct distribution: the true distribution has heavy-tailed or asymmetric distribution, the proposed procedures also work well except for a small sample size and a small causal effects situation. 2SRI also work well when using model selection procedures in large sample situation; this is the different result from Basu et al. (2017).

As we mentioned, the proposed procedures are significant contributions in the situation where there are some unmeasured covariates and nonlinear outcomes, since there has been no research that considered model selection procedures when we need to specify both the true treatment model and the true outcome model. However, there are some future works. At first, a binary outcome is only considered in this article. Since LIML is full-likelihood approach, the method can be expanded to more complex models, for instance, more general outcome of an exponential family, or a time-to-event outcome (Kianian et al., 2019 and Mart\'{i}nez-Camblor et al., 2019). Our proposed method need to consider both an outcome and a treatment likelihood, however, the other restrictions are limited. For instance, our proposed method is not restricted to a binary instrumental variable (Wang and Tchetgen Tchetgen, 2018 and Kianian et al., 2019) and continuous treatment (Mart\'{i}nez-Camblor et al., 2019) situation. Therefore, our proposed procedure has wide range of expandability. Next, an impact of misspecification of an unmeasured covariate distribution needs to be confirmed continuously. As our simulation results, the impact is only limited; however, the behavior of estimates in the other situations is not clear. Especially, we cannot conclude the result of under an asymmetric distribution through our simulation setting only. It is necessary to continue the simulation-based considerations.


\newpage
\appendix
\section{Regularity conditions}
{\bf C.1} to {\bf C.6} are needed for the derivation of the LAIC and the proof of the LBIC.
\begin{description}
\item[C.1:]
$$
{\rm E}_{\bld{\theta}^{0}}\left[\left(\frac{\partial}{\partial\bld{\theta}}\ell(\bld{\theta})\right)^{\otimes 2}\right]>O.
$$
\item[C.2:] $\Theta$ is a compact set.
\item[C.3:] $\bld{\theta}^{0}\in interior(\Theta)$, where $\bld{\theta}^{0}$ is the true value of $\bld{\theta}$.
\item[C.4:] 
$$
\sum_{y}\int\sup_{\bld{\theta}\in \mathcal{N}}\left|\left|\frac{\partial}{\partial\bld{\theta}}f(y,w|\bld{z},\bld{x};\bld{\theta})\right|\right|dw<\infty,\ \ \sum_{y}\int\sup_{\bld{\theta}\in \mathcal{N}}\left|\left|\frac{\partial^2}{\partial\bld{\theta}^{\otimes 2}}f(y,w|\bld{z},\bld{x};\bld{\theta})\right|\right|dw<\infty,
$$
where $\mathcal{N}$ is a neighborhood of $\bld{\theta}^{0}$.
\item[C.5:]
$$
\sum_{y}\int\sup_{\bld{\theta}\in \mathcal{N}}\left|\left|\frac{\partial^2}{\partial\bld{\theta}^{\otimes 2}}\ell(\bld{\theta})\right|\right|dw<\infty.
$$
\end{description}
Note that {\bf C.1} is an assumption for non-diagonal elements, since diagonal elements of the matrix are clearly positive.
\begin{description}
\item[C.6:] For all $i,j,k$,
$$
\sum_{y}\int\sup_{\bld{\theta}\in \mathcal{N}}\left|\frac{\partial^3}{\partial\theta_{i}\partial\theta_{j}\partial\theta_{k}}\ell(\bld{\theta})\right|dw<\infty.
$$
\end{description}
\section{Calculations, derivation, and proof}
\subsection{Calculation of (\ref{for1_5})}
When $y=1$,
\begin{align}
\label{ap1_1}
P(Y=1|w,\bld{z},\bld{x};\bld{\theta})&={\rm E}_{\bld{\theta}}\left[Y|w,\bld{x},\bld{z}\right]={\rm E}_{\bld{\theta}}\left[\bld{1}\left\{\varphi_{2}\left(w,\bld{x};\bld{\beta}\right)+U\geq 0\right\}|w,\bld{x},\bld{z}\right]\nonumber\\
&=\int\bld{1}\left\{u\geq -\varphi_{2}\left(w,\bld{x};\bld{\beta}\right)\right\}f(u|w,\bld{x},\bld{z};\bld{\theta})du\nonumber\\
&=\int\bld{1}\left\{u\geq -\varphi_{2}\left(w,\bld{x};\bld{\beta}\right)\right\}f(u|v,\bld{x},\bld{z};\bld{\theta})du\nonumber\\
&=\int\bld{1}\left\{\omega\geq -\frac{\varphi_{2}\left(w,\bld{x};\bld{\beta}\right)+\rho v(\bld{\alpha})}{\sqrt{1-\rho^2}}\right\}\phi(\omega)d\omega\nonumber\\
&=\Phi\left(\frac{\varphi_{2}\left(w,\bld{x};\bld{\beta}\right)+\rho v(\bld{\alpha})}{\sqrt{1-\rho^2}}\right).
\end{align}
Note that $v(\bld{\alpha})=w-\varphi_{1}(\bld{z},\bld{x};\bld{\alpha})$ and $U|v\sim N\left(\rho v,1-\rho^2\right)$. When $y=0$, the flow of the calculation is the same as that of (\ref{ap1_1}).

\subsection{Calculation of (\ref{for1_5_2})}
We only consider when $y_{i}=1$ and $w_{i}=1$. When the other situations, log-likelihood can be derived in the same manner.
\begin{align}
\label{ap2_1}
\log\left\{{\rm P}\left(y_{i}=1,w_{i}=1|\bld{z}_{i},\bld{x}_{i};\bld{\theta}\right)^{y_{i}w_{i}}\right\}&=y_{i}w_{i}\log\left\{\int^{\infty}_{-\varphi_{i2}(\bld{\beta})}\int^{\infty}_{-\varphi_{i1}(\bld{\alpha})}f(v,u;\xi)dvdu\right\},
\end{align}
where $f(v,u;\xi)$ is a joint density of $(V,U)$. Regarding $\log\{\cdot\}$ of (\ref{ap2_1}), 
$$
\int^{\infty}_{-\varphi_{i2}(\bld{\beta})}\int^{\infty}_{-\varphi_{i1}(\bld{\alpha})}f(v,u;\xi)dvdu=1-F(\infty,-\varphi_{i2}(\bld{\beta});\xi)-F(-\varphi_{i1}(\bld{\alpha}),\infty;\xi)+F(-\varphi_{i1}(\bld{\alpha}),-\varphi_{i2}(\bld{\beta});\xi).
$$

\subsection{Derivation of the LAIC}
The flow of the derivation is the same as that of Taguri et al. (2014). Note that regularity conditions {\bf C.1} to {\bf C.6} need to be assumed. To confirm the bias of an empirical log-likelihood, we need to conduct a Taylor expansion around $\bld{\theta}^{0}$.
\begin{align}
\label{B4_1}
-2\ell(\bld{\theta})=-2\ell(\bld{\theta}^{0})-2\frac{\partial\ell(\bld{\theta}^{0})}{\partial\bld{\theta}^{\top}}(\bld{\theta}-\bld{\theta}^{0})-(\bld{\theta}-\bld{\theta}^{0})^{\top}\frac{\partial^2\ell(\bld{\theta}^{0})}{\partial\bld{\theta}^{\otimes 2}}(\bld{\theta}-\bld{\theta}^{0})+o_{p}(1).
\end{align}
From (\ref{B4_1}),
\begin{align}
\label{B4_2}
-2d_{n}(\hat{\bld{\theta}})&=-2{\rm E}\left[\ell(\bld{\theta}^{0})\right]-2{\rm E}\left[\frac{\partial\ell(\bld{\theta}^{0})}{\partial\bld{\theta}^{\top}}\right](\hat{\bld{\theta}}-\bld{\theta}^{0})-(\hat{\bld{\theta}}-\bld{\theta}^{0})^{\top}{\rm E}\left[\frac{\partial^2\ell(\bld{\theta}^{0})}{\partial\bld{\theta}^{\otimes 2}}\right](\hat{\bld{\theta}}-\bld{\theta}^{0})+o_{p}(1)\nonumber\\
&=-2{\rm E}\left[\ell(\bld{\theta}^{0})\right]-n(\hat{\bld{\theta}}-\bld{\theta}^{0})^{\top}{\rm E}\left[\frac{\partial^2\ell(\bld{\theta}^{0})}{\partial\bld{\theta}^{\otimes 2}}\right](\hat{\bld{\theta}}-\bld{\theta}^{0})+o_{p}(1).
\end{align}
Note that the expectation of the true score function is $\bld{0}$ under the true parameter $\bld{\theta}^{0}$. Also, using the equation
\begin{align}
\label{C1_3}
\bld{0}=\frac{\partial\ell(\hat{\bld{\theta}})}{\partial\bld{\theta}}=\frac{\partial\ell(\bld{\theta}^{0})}{\partial\bld{\theta}}+\frac{\partial\ell(\bld{\theta}^{0})}{\partial\bld{\theta}^{\top}}(\hat{\bld{\theta}}-\bld{\theta}^{0})+O_{p}(1),
\end{align}
\begin{align}
\label{B4_3}
-2\hat{D}_{n}(\hat{\bld{\theta}})&=-2\ell(\bld{\theta}^{0})-2\frac{\partial\ell(\bld{\theta}^{0})}{\partial\bld{\theta}^{\top}}(\hat{\bld{\theta}}-\bld{\theta}^{0})-(\hat{\bld{\theta}}-\bld{\theta}^{0})^{\top}\frac{\partial^2\ell(\bld{\theta}^{0})}{\partial\bld{\theta}^{\otimes 2}}(\hat{\bld{\theta}}-\bld{\theta}^{0})+o_{p}(1)\nonumber\\
&=-2\ell(\bld{\theta}^{0})+(\hat{\bld{\theta}}-\bld{\theta}^{0})^{\top}\frac{\partial^2\ell(\bld{\theta}^{0})}{\partial\bld{\theta}^{\otimes 2}}(\hat{\bld{\theta}}-\bld{\theta}^{0})+o_{p}(1)\nonumber\\
&=-2\ell(\bld{\theta}^{0})+n(\hat{\bld{\theta}}-\bld{\theta}^{0})^{\top}{\rm E}\left[\frac{\partial^2\ell(\bld{\theta}^{0})}{\partial\bld{\theta}^{\otimes 2}}\right](\hat{\bld{\theta}}-\bld{\theta}^{0})+o_{p}(1).
\end{align}
From (\ref{B4_2}) and (\ref{B4_3}), the expected bias between the expected and the empirical log-likelihood is
\begin{align*}
-2n{\rm E}\left[d_{n}(\hat{\bld{\theta}})-\hat{D}_{n}(\hat{\bld{\theta}})\right]&=-2n{\rm E}\left[(\hat{\bld{\theta}}-\bld{\theta}^{0})^{\top}{\rm E}\left[\frac{\partial^2\ell(\bld{\theta}^{0})}{\partial\bld{\theta}^{\otimes 2}}\right](\hat{\bld{\theta}}-\bld{\theta}^{0})\right]+o(1)\\
&=-2tr\left\{{\rm E}\left[\frac{\partial^2\ell(\bld{\theta}^{0})}{\partial\bld{\theta}^{\otimes 2}}\right]\times n{\rm E}\left[(\hat{\bld{\theta}}-\bld{\theta}^{0})^{\otimes 2}\right]\right\}+o(1)\\
&\approx2|\hat{\bld{\theta}}|.
\end{align*}
Therefore, the LAIC is shown to have the required properties.
\subsection{Proof of Theorem \ref{theo2}}
The proof is based on Noghrehchi et al. (2021), the same flow as ordinary BIC. Note that regularity conditions {\bf C.1} to {\bf C.6} need to be assumed. Before the main proof, we calculate a difference of LBIC values between pairs of models.
\begin{align}
\label{C1_1}
LBIC\left(m_{\mathcal{A}}^{0},m_{\mathcal{B}}^{0}\right)-LBIC\left(m_{\mathcal{A}},m_{\mathcal{B}}\right)&=-2\ell\left(\hat{\bld{\alpha}}_{m^{0}_{\mathcal{A}}},\hat{\bld{\beta}}_{m^{0}_{\mathcal{B}}}\right)+\left(\left|\hat{\bld{\alpha}}_{m^{0}_{\mathcal{A}}}\right|+\left|\hat{\bld{\beta}}_{m^{0}_{\mathcal{B}}}\right|\right)\log(n)\nonumber\\
&\hspace{0.5cm}-\left(-2\ell\left(\hat{\bld{\alpha}}_{m_{\mathcal{A}}},\hat{\bld{\beta}}_{m_{\mathcal{B}}}\right)+\left(\left|\hat{\bld{\alpha}}_{m_{\mathcal{A}}}\right|+\left|\hat{\bld{\beta}}_{m_{\mathcal{B}}}\right|\right)\log(n)\right)\nonumber\\
&=-2\log\left(\frac{L\left(\hat{\bld{\alpha}}_{m^{0}_{\mathcal{A}}},\hat{\bld{\beta}}_{m^{0}_{\mathcal{B}}}\right)}{L\left(\hat{\bld{\alpha}}_{m_{\mathcal{A}}},\hat{\bld{\beta}}_{m_{\mathcal{B}}}\right)}\right)\nonumber\\
&\hspace{0.5cm}+\left(\left|\hat{\bld{\alpha}}_{m^{0}_{\mathcal{A}}}\right|-\left|\hat{\bld{\alpha}}_{m_{\mathcal{A}}}\right|+\left|\hat{\bld{\beta}}_{m^{0}_{\mathcal{B}}}\right|-\left|\hat{\bld{\beta}}_{m_{\mathcal{B}}}\right|\right)\log(n).\nonumber\\
\end{align}
We will proceed with the proof by dividing it into the following two situations:
\begin{enumerate}
\item $m_{\mathcal{A}}\in\mathcal{A}^{c}\backslash \left\{m_{\mathcal{A}}^{0}\right\},\, m_{\mathcal{B}}\in\mathcal{B}^{c}\backslash \left\{m_{\mathcal{B}}^{0}\right\}$
\item $\left(m_{\mathcal{A}}\in\mathcal{A}\backslash \mathcal{A}^{c},\, m_{\mathcal{B}}\in\mathcal{B}^{c}\backslash \left\{m_{\mathcal{B}}^{0}\right\}\right)\ \ or\ \ \left(m_{\mathcal{A}}\in\mathcal{A}^{c}\backslash \left\{m_{\mathcal{A}}^{0}\right\},\, m_{\mathcal{B}}\in\mathcal{B}\backslash \mathcal{B}^{c}\right)\\
or\ \ \left(m_{\mathcal{A}}\in\mathcal{A}\backslash \mathcal{A}^{c},\, m_{\mathcal{B}}\in\mathcal{B}\backslash \mathcal{B}^{c}\right)$.
\end{enumerate}
Note that the above two situations are mutually exclusive.

First, we will confirm consistency for situation 1. Since $m_{\mathcal{A}}^{0},\, m_{\mathcal{B}}^{0}$ are the smallest models in $\mathcal{A}^{c},\, \mathcal{B}^{c}$, the second term of (\ref{C1_1}) becomes negative from (2.11). Subsequently, to confirmthe first term of (\ref{C1_1}), i.e., the likelihood ratio term, we conduct a Taylor expansion around $\hat{\bld{\theta}}$:
\begin{align*}
-2\ell(\bld{\theta}^{0})&=-2\ell(\hat{\bld{\theta}})-2\frac{\partial\ell(\hat{\bld{\theta}})}{\partial\bld{\theta}^{\top}}(\bld{\theta}^{0}-\hat{\bld{\theta}})-(\bld{\theta}^{0}-\hat{\bld{\theta}})^{\top}\frac{\partial^2\ell(\hat{\bld{\theta}})}{\partial\bld{\theta}^{\otimes 2}}(\bld{\theta}^{0}-\hat{\bld{\theta}})+o_{p}(1)\\
&=-2\ell(\hat{\bld{\theta}})-(\bld{\theta}^{0}-\hat{\bld{\theta}})^{\top}\frac{\partial^2\ell(\hat{\bld{\theta}})}{\partial\bld{\theta}^{\otimes 2}}(\bld{\theta}^{0}-\hat{\bld{\theta}})+o_{p}(1)\\
&=-2\ell(\hat{\bld{\theta}})-n(\bld{\theta}^{0}-\hat{\bld{\theta}})^{\top}{\rm E}\left[\frac{\partial^2\ell(\bld{\theta}^{0})}{\partial\bld{\theta}^{\otimes 2}}\right](\bld{\theta}^{0}-\hat{\bld{\theta}})+o_{p}(1).
\end{align*}
Note that the empirical score function is $\bld{0}$ for an estimated parameter $\hat{\bld{\theta}}$. Therefore,
\begin{align}
\label{C1_2}
-2\ell(\hat{\bld{\theta}})=-2\ell(\bld{\theta}^{0})+n(\bld{\theta}^{0}-\hat{\bld{\theta}})^{\top}{\rm E}\left[\frac{\partial^2\ell(\bld{\theta}^{0})}{\partial\bld{\theta}^{\otimes 2}}\right](\bld{\theta}^{0}-\hat{\bld{\theta}})+o_{p}(1).
\end{align}
Also, by using (\ref{C1_3}), (\ref{C1_2}) becomes
\begin{align}
\label{C1_4}
-2\ell(\hat{\bld{\theta}})&=-2\ell(\bld{\theta}^{0})+\frac{1}{\sqrt{n}}\frac{\partial\ell(\bld{\theta}^{0})}{\partial\bld{\theta}^{\top}}\left({\rm E}\left[\frac{\partial^2\ell(\bld{\theta}^{0})}{\partial\bld{\theta}^{\otimes 2}}\right]\right)^{-1}\frac{1}{\sqrt{n}}\frac{\partial\ell(\bld{\theta}^{0})}{\partial\bld{\theta}}+o_{p}(1)\nonumber\\
&=-2\ell(\bld{\theta}^{0})-S^{\top}(\bld{\theta}^{0})\mathcal{I}^{-1}(\bld{\theta}^{0})S(\bld{\theta}^{0})+o_{p}(1),
\end{align}
where $S(\bld{\theta}^{0})$ and $\mathcal{I}(\bld{\theta}^{0})$ are the score function and the Fisher information matrix, respectively. Regarding the minimum model, (\ref{C1_4}) becomes
\begin{align}
\label{C1_5}
-2\ell(\hat{\bld{\theta}}_{0})=-2\ell(\bld{\theta}^{0}_{0})-S^{\top}(\bld{\theta}^{0}_{0})\mathcal{I}^{-1}(\bld{\theta}^{0}_{0})S(\bld{\theta}^{0}_{0})+o_{p}(1).
\end{align}
Note that $\bld{\theta}^{0}_{0}$ represents the true parameter of the minimum model, and is represented as
$$
\bld{\theta}^{0}=\left(
\begin{array}{c}
\bld{\theta}^{0}_{0}\\
\bld{0}
\end{array}
\right).
$$
Therefore, $-2\ell(\bld{\theta}^{0})=-2\ell(\bld{\theta}^{0}_{0})$. Continuing, we define $G=\left(I,O\right)^{\top}$ and use this definition to represent $S(\bld{\theta}^{0}_{0})$ and $\mathcal{I}(\bld{\theta}^{0})$ as follows:
$$
S(\bld{\theta}^{0}_{0})=G^{\top}S(\bld{\theta}^{0}),\ \ \mathcal{I}(\bld{\theta}^{0}_{0})=G^{\top}\mathcal{I}(\bld{\theta}^{0})G
$$
respectively. From (\ref{C1_4}) and (\ref{C1_5}),
\begin{align}
\label{C1_6}
-2\log\left(\frac{L\left(\hat{\bld{\alpha}}_{m^{0}_{\mathcal{A}}},\hat{\bld{\beta}}_{m^{0}_{\mathcal{B}}}\right)}{L\left(\hat{\bld{\alpha}}_{m_{\mathcal{A}}},\hat{\bld{\beta}}_{m_{\mathcal{B}}}\right)}\right)&=-2\ell(\hat{\bld{\theta}}_{0})-(-2\ell(\hat{\bld{\theta}}))\nonumber\\
&=S^{\top}(\bld{\theta}^{0})\mathcal{I}^{-1}(\bld{\theta}^{0})S(\bld{\theta}^{0})-S^{\top}(\bld{\theta}^{0}_{0})\mathcal{I}^{-1}(\bld{\theta}^{0}_{0})S(\bld{\theta}^{0}_{0})+o_{p}(1)\nonumber\\
&=S^{\top}(\bld{\theta}^{0})\left(\mathcal{I}^{\frac{1}{2}}(\bld{\theta}^{0})\right)^{-1}\underline{\left(I-\mathcal{I}^{\frac{1}{2}}(\bld{\theta}^{0})G\mathcal{I}^{-1}(\bld{\theta}^{0}_{0})G^{\top}\mathcal{I}^{\frac{1}{2}}(\bld{\theta}^{0})\right)}\left(\mathcal{I}^{\frac{1}{2}}(\bld{\theta}^{0})\right)^{-1}S(\bld{\theta}^{0}),\nonumber\\
\end{align}
where $\mathcal{I}^{\frac{1}{2}}(\bld{\theta}^{0})$ is a lower triangular matrix; i.e.,
$$
\mathcal{I}(\bld{\theta}^{0})=\mathcal{I}^{\frac{1}{2}}(\bld{\theta}^{0})\left(\mathcal{I}^{\frac{1}{2}}(\bld{\theta}^{0})\right)^{\top}.
$$
Regarding the score function,
$$
\left(\mathcal{I}^{\frac{1}{2}}(\bld{\theta}^{0})\right)^{-1}S(\bld{\theta}^{0})\stackrel{L}{\to}N(\bld{0},I),
$$
and the underlined part of (\ref{C1_6}) satisfies the following:
\begin{align*}
\left(I-\mathcal{I}^{\frac{1}{2}}(\bld{\theta}^{0})G\mathcal{I}^{-1}(\bld{\theta}^{0}_{0})G^{\top}\mathcal{I}^{\frac{1}{2}}(\bld{\theta}^{0})\right)\left(I-\mathcal{I}^{\frac{1}{2}}(\bld{\theta}^{0})G\mathcal{I}^{-1}(\bld{\theta}^{0}_{0})G^{\top}\mathcal{I}^{\frac{1}{2}}(\bld{\theta}^{0})\right)&\\
&\hspace{-11.5cm}=I-2\mathcal{I}^{\frac{1}{2}}(\bld{\theta}^{0})G\mathcal{I}^{-1}(\bld{\theta}^{0}_{0})G^{\top}\mathcal{I}^{\frac{1}{2}}(\bld{\theta}^{0})+\mathcal{I}^{\frac{1}{2}}(\bld{\theta}^{0})G\mathcal{I}^{-1}(\bld{\theta}^{0}_{0})G^{\top}\mathcal{I}(\bld{\theta}^{0})G\mathcal{I}^{-1}(\bld{\theta}^{0}_{0})G^{\top}\mathcal{I}^{\frac{1}{2}}(\bld{\theta}^{0})\\
&\hspace{-11.5cm}=I-\mathcal{I}^{\frac{1}{2}}(\bld{\theta}^{0})G\mathcal{I}^{-1}(\bld{\theta}^{0}_{0})G^{\top}\mathcal{I}^{\frac{1}{2}}(\bld{\theta}^{0}).
\end{align*}
Therefore, from statement (ii) of Rao (1973, P.186), (\ref{C1_6}) becomes
\begin{align}
\label{C1_7}
-2\log\left(\frac{L\left(\hat{\bld{\alpha}}_{m^{0}_{\mathcal{A}}},\hat{\bld{\beta}}_{m^{0}_{\mathcal{B}}}\right)}{L\left(\hat{\bld{\alpha}}_{m_{\mathcal{A}}},\hat{\bld{\beta}}_{m_{\mathcal{B}}}\right)}\right)\stackrel{L}{\to}\Lambda\sim\chi^{2}_{|\hat{\bld{\theta}}|-|\hat{\bld{\theta}}_{0}|}.
\end{align}
From (2.11) and (\ref{C1_7}),
\begin{align}
\label{C1_8}
{\rm P}\left(LBIC\left(m_{\mathcal{A}}^{0},m_{\mathcal{B}}^{0}\right)-LBIC\left(m_{\mathcal{A}},m_{\mathcal{B}}\right)<0\right)\nonumber\\
&\hspace{-5.5cm}={\rm P}\left(-2\log\left(\frac{L\left(\hat{\bld{\alpha}}_{m^{0}_{\mathcal{A}}},\hat{\bld{\beta}}_{m^{0}_{\mathcal{B}}}\right)}{L\left(\hat{\bld{\alpha}}_{m_{\mathcal{A}}},\hat{\bld{\beta}}_{m_{\mathcal{B}}}\right)}\right)<\left(|\hat{\bld{\theta}}|-|\hat{\bld{\theta}}_{0}|\right)\log(n)\right)\to1.
\end{align}

Next, we will confirm consistency for situation 2. From (\ref{C1_1}),
\begin{align*}
\frac{1}{2}\left(LBIC\left(m_{\mathcal{A}},m_{\mathcal{B}}\right)-LBIC\left(m_{\mathcal{A}}^{0},m_{\mathcal{B}}^{0}\right)\right)&=\ell(\hat{\bld{\theta}}_{0})-\ell(\hat{\bld{\theta}})+\frac{1}{2}\left(|\hat{\bld{\theta}}|-|\hat{\bld{\theta}}_{0}|\right)\log(n)\\
&=\ell(\hat{\bld{\theta}}_{0})-\ell(\hat{\bld{\theta}})+O(\log(\sqrt{n}))
\end{align*}
From (\ref{C1_4}),
\begin{align*}
\ell(\hat{\bld{\theta}}_{0})-\ell(\hat{\bld{\theta}})&=\ell(\bld{\theta}^{0}_{0})+\frac{1}{2}S^{\top}(\bld{\theta}^{0}_{0})\mathcal{I}^{-1}(\bld{\theta}^{0}_{0})S(\bld{\theta}^{0}_{0})-\ell(\bld{\theta}^{0})-\frac{1}{2}S^{\top}(\bld{\theta}^{0})\mathcal{I}^{-1}(\bld{\theta}^{0})S(\bld{\theta}^{0})+o_{p}(1)\\
&=\ell(\bld{\theta}^{0}_{0})-\ell(\bld{\theta}^{0})+\frac{1}{2}\left(S^{\top}(\bld{\theta}^{0}_{0})\mathcal{I}^{-1}(\bld{\theta}^{0}_{0})S(\bld{\theta}^{0}_{0})-S^{\top}(\bld{\theta}^{0})\mathcal{I}^{-1}(\bld{\theta}^{0})S(\bld{\theta}^{0})\right)+o_{p}(1)\\
&=O_{p}(n)\pm O_{p}(1)+o_{p}(1)\\
&=O_{p}(n).
\end{align*}
Therefore, model selection consistency is proved.

\section{Supplementary information for simulations}
Data generating programs, simulation datasets, simulation programs, simulation results, and programs for deriving tables and figures are available from the following URL: 
\begin{itemize}
\item \url{https://www.hogehoge}
\end{itemize}
\setcounter{table}{0}
\renewcommand{\thetable}{C\arabic{table}}
\subsection{Continuous and linear treatment effect}
\subsubsection{Candidates of models}
To select a treatment model and an outcome model, we present the following respective candidate models.
\begin{description}
\item{{\bf Candidates for treatment models}}
$$
W=\alpha_{0}+z\alpha_{1}+x_{2}\alpha_{2}+x_{3}\alpha_{3}+\left(z\times x_{2}\right)\alpha_{4}+\left(z\times x_{3}\right)\alpha_{5}+\left(x_{2}\times x_{3}\right)\alpha_{6}+V
$$

\begin{table}[h]
\begin{center}
\caption{Settings of candidates for treatment models}
\begin{tabular}{|c|c|c|c|c|c|c|c|}\hline
{{\bf Candidate model}}&$\alpha_{0}$&$\alpha_{1}$&$\alpha_{2}$&$\alpha_{3}$&$\alpha_{4}$&$\alpha_{5}$&$\alpha_{6}$\\\hline\hline
Model a1&$\neq 0$&$\neq0$&$=0$&$=0$&$=0$&$=0$&$=0$\\\hline
Model a2&$\neq 0$&$\neq0$&$\neq0$&$=0$&$=0$&$=0$&$=0$\\\hline
Model a3&$\neq 0$&$\neq0$&$=0$&$\neq0$&$=0$&$=0$&$=0$\\\hline
Model a4&$\neq 0$&$\neq0$&$\neq0$&$\neq0$&$=0$&$=0$&$=0$\\\hline
Model a5&$\neq 0$&$\neq0$&$\neq0$&$=0$&$\neq0$&$=0$&$=0$\\\hline
Model a6&$\neq 0$&$\neq0$&$=0$&$\neq0$&$=0$&$\neq0$&$=0$\\\hline
Model a7&$\neq 0$&$\neq0$&$\neq0$&$\neq0$&$\neq0$&$\neq0$&$\neq0$\\\hline
\end{tabular}
\end{center}
\end{table}
Note that Model a4 is the true model. From Model a1 to Model a3, Model a5, and Model a6 are misspecified models.
\item{{\bf Candidates for outcome models}}
\begin{align*}
Y=\bld{1}\left\{\beta_{0}+w\beta_{1}+x_{1}\beta_{2}+x_{2}\beta_{3}+x_{3}\beta_{4}+\left(x_{1}\times x_{2}\right)\beta_{5}+\left(x_{1}\times x_{3}\right)\beta_{6}+\left(x_{2}\times x_{3}\right)\beta_{7}+U\geq0\right\}
\end{align*}

\begin{table}[h]
\begin{center}
\caption{Settings of candidates for outcome models}
\begin{tabular}{|c|c|c|c|c|c|c|c|c|}\hline
{{\bf Candidate model}}&$\beta_{0}$&$\beta_{1}$&$\beta_{2}$&$\beta_{3}$&$\beta_{4}$&$\beta_{5}$&$\beta_{6}$&$\beta_{7}$\\\hline\hline
Model b1&$\neq 0$&$\neq0$&$=0$&$=0$&$=0$&$=0$&$=0$&$=0$\\\hline
Model b2&$\neq 0$&$\neq0$&$\neq0$&$\neq0$&$=0$&$=0$&$=0$&$=0$\\\hline
Model b3&$\neq 0$&$\neq0$&$\neq0$&$\neq0$&$\neq0$&$=0$&$=0$&$=0$\\\hline
Model b4&$\neq 0$&$\neq0$&$\neq0$&$\neq0$&$=0$&$\neq0$&$=0$&$=0$\\\hline
Model b5&$\neq 0$&$\neq0$&$\neq0$&$\neq0$&$\neq0$&$\neq0$&$\neq0$&$\neq0$\\\hline
\end{tabular}
\end{center}
\end{table}
Note that Model b2 is the true model. Model b1 is a misspecified model. For 2SRI, a residual term is also included in the model.
\end{description}
\subsubsection{Boxplots of descriptive statistics for each estimator (continuous and linear treatment effect)}
\newpage
\newpage
\newpage
\subsection{Dichotomous and heterogeneous treatment effect}
\subsubsection{Candidates of models}
\begin{description}
\item{{\bf Candidates for outcome models}}
\begin{align*}
Y&=\bld{1}\left\{\beta_{0}+w\beta_{1}+x_{1}\beta_{2}+x_{2}\beta_{3}+x_{3}\beta_{4}+\left(w\times x_{1}\right)\beta_{5}\right.\\
&\hspace{0.5cm}\left.+\left(x_{1}\times x_{2}\right)\beta_{6}+\left(x_{1}\times x_{3}\right)\beta_{7}+\left(x_{2}\times x_{3}\right)\beta_{8}+U\geq0\right\}
\end{align*}

\begin{table}[h]
\begin{center}
\caption{Settings of candidates for outcome models}
\begin{tabular}{|c|c|c|c|c|c|c|c|c|c|}\hline
{{\bf Candidate model}}&$\beta_{0}$&$\beta_{1}$&$\beta_{2}$&$\beta_{3}$&$\beta_{4}$&$\beta_{5}$&$\beta_{6}$&$\beta_{7}$&$\beta_{8}$\\\hline\hline
Model b1&$\neq 0$&$\neq0$&$=0$&$=0$&$=0$&$=0$&$=0$&$=0$&$=0$\\\hline
Model b2&$\neq 0$&$\neq0$&$\neq0$&$\neq0$&$=0$&$=0$&$=0$&$=0$&$=0$\\\hline
Model b3&$\neq 0$&$\neq0$&$\neq0$&$\neq0$&$\neq0$&$=0$&$=0$&$=0$&$=0$\\\hline
Model b4&$\neq 0$&$\neq0$&$\neq0$&$\neq0$&$\neq0$&$=0$&$\neq0$&$\neq0$&$\neq0$\\\hline
Model b5&$\neq 0$&$\neq0$&$\neq0$&$\neq0$&$=0$&$\neq0$&$=0$&$=0$&$=0$\\\hline
Model b6&$\neq 0$&$\neq0$&$\neq0$&$\neq0$&$\neq0$&$\neq0$&$=0$&$=0$&$=0$\\\hline
Model b7&$\neq 0$&$\neq0$&$\neq0$&$\neq0$&$\neq0$&$\neq0$&$\neq0$&$\neq0$&$\neq0$\\\hline
\end{tabular}
\end{center}
\end{table}
Note that Model b5 is the true model, and from Model b1 to Model b4 are misspecified models.
\end{description}

\subsubsection{Boxplots of descriptive statistics for each estimator (dichotomous and heterogeneous)}
\newpage
\newpage
\newpage

\newpage
\newpage
\newpage

\newpage
\newpage
\newpage

\newpage
\begin{landscape}
\begin{figure}[h]
\begin{center}
\begin{tabular}{c}
\includegraphics[width=15cm]{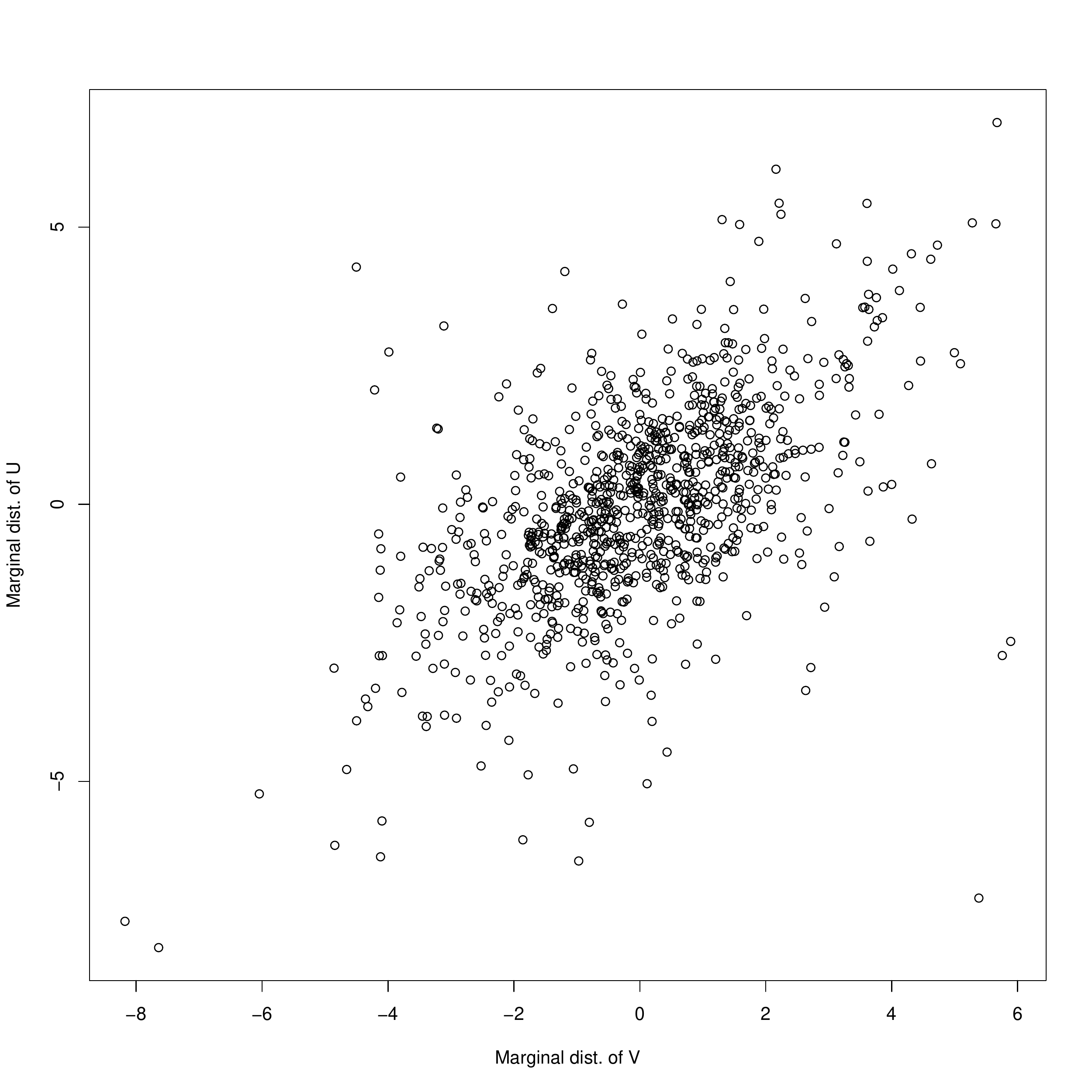}
\end{tabular}
\caption{One example of unmeasured covariates distribution (strong correlation situation, logistic reg. \& t-copula)}
\label{fig9}
\end{center}
\end{figure}
\end{landscape}\newpage

\newpage
\begin{landscape}
\begin{figure}[h]
\begin{center}
\begin{tabular}{c}
\includegraphics[width=15cm]{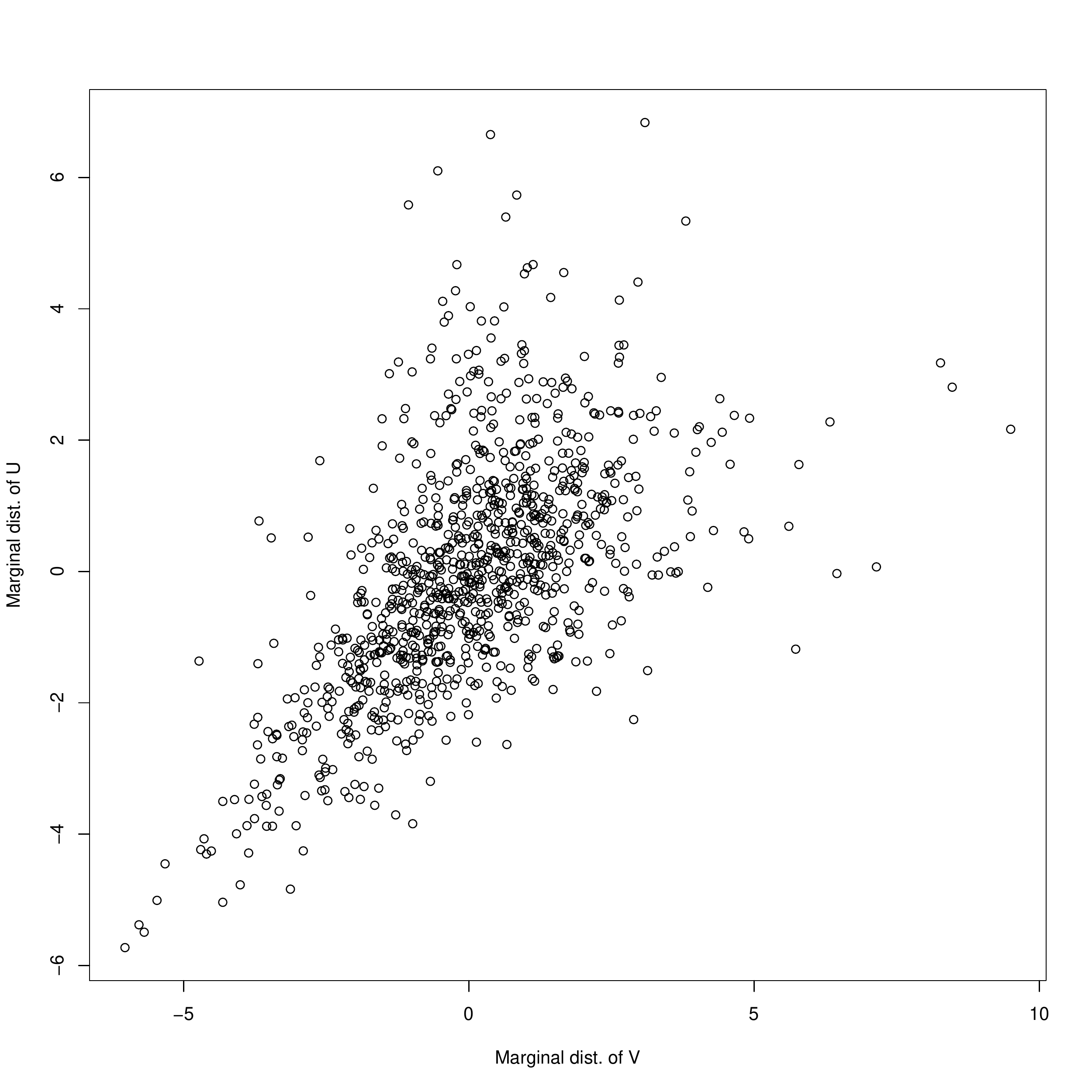}
\end{tabular}
\caption{One example of unmeasured covariates distribution (strong correlation situation, logistic reg. \& Clayton-copula) 2/2}
\label{fig10}
\end{center}
\end{figure}
\end{landscape}\newpage

\end{document}